\documentclass[11pt]{article}

\usepackage{amssymb}
\usepackage{citesort}
\usepackage{amsmath,mathrsfs}
\usepackage{url,graphicx}
\usepackage{upgreek}

\def\si{\sigma}
\def\de{\delta}

\def\na{\nabla}
\def\ka{\kappa}

\def\pa{\partial}

\def\al{\alpha}

\def\ii{\textrm i}
\def\ee{\textrm e}
\def\ca{{\boldsymbol{\mathcal{A}}}}
\def\ue{\upeta}
\def\ua{\upalpha}
\def\vp{\varphi}

\def\ud{\textrm{d}}
\def\bn{{\boldsymbol \nabla}}

\newcommand{\dbb}{de$\,$Broglie-Bohm}

\newcommand{\be}{\begin{equation}}
\newcommand{\en}{\end{equation}}
\newcommand{\bea}{\begin{eqnarray}}
\newcommand{\ena}{\end{eqnarray}}

\title{Semi-classical approximations based on Bohmian mechanics}

\author{ Ward Struyve\footnote{Instituut voor Theoretische Fysica \& Centrum voor Logica en Filosofie van de Wetenschappen, KU Leuven, Belgium. E-mail: ward.struyve@gmail.com}
}

\date{}

\begin{document}

\maketitle

\begin{abstract}
\noindent 
Semi-classical theories are approximations to quantum theory that treat some degrees of freedom classically and others quantum mechanically. In the usual approach, the quantum degrees of freedom are described by a wave function which evolves according to some Schr\"odinger equation with a Hamiltonian that depends on the classical degrees of freedom. The classical degrees of freedom satisfy classical equations that depend on the expectation values of quantum operators. In this paper, we study an alternative approach based on Bohmian mechanics. In Bohmian mechanics the quantum system is not only described by the wave function, but also with additional variables such as particle positions or fields. By letting the classical equations of motion depend on these variables, rather than the quantum expectation values, a semi-classical approximation is obtained that is closer to the exact quantum results than the usual approach. We discuss the Bohmian semi-classical approximation in various contexts, such as non-relativistic quantum mechanics, quantum electrodynamics and quantum gravity. The main motivation comes from quantum gravity. The quest for a quantum theory for gravity is still going on. Therefore a semi-classical approach where gravity is treated classically may be an approximation that already captures some quantum gravitational aspects. The Bohmian semi-classical theories will be derived from the full Bohmian theories. In the case there are gauge symmetries, like in quantum electrodynamics or quantum gravity, special care is required. In order to derive a consistent semi-classical theory it will be necessary to isolate gauge-independent dependent degrees of freedom from gauge degrees of freedom and consider the approximation where some of the former are considered classical.

\end{abstract}

\bibliographystyle{unsrt,plain}

\section{Introduction}
Quantum gravity is often considered to be the holy grail of theoretical physics. One approach is canonical quantum gravity, which concerns the Wheeler-DeWitt equation and which is obtained by applying the usual quantization methods (which were so successful in the case of high energy physics) to Einstein's field equations. However, this approach suffers from a host of problems, some of technical and some of conceptual nature (such as finding solutions to the Wheeler-DeWitt equation, the problem of time, \dots). For this reason one often resorts to a semi-classical approximation where gravity is treated classically and matter quantum mechanically \cite{wald94,kiefer04}. The hope is that such an approximation is easier to analyze and yet reveals some effects of quantum gravitational nature.

In the usual approach to semi-classical gravity, matter is described by quantum field theory on curved space-time. For example, in the case the matter is described by a quantized scalar field, the state vector can be considered a functional $\Psi(\vp)$ on the space of fields, which satisfies a particular Schr\"odinger equation
\be
\ii \pa_t \Psi(\vp,t) = {\widehat H}(\vp,g) \Psi(\vp,t) \,,
\label{0.001}
\en
where the Hamiltonian operator ${\widehat H}$ depends on the classical space-time metric $g$. This metric satisfies Einstein's field equations
\be
G_{\mu \nu} (g) = 8\pi G \langle \Psi | {\widehat T}_{\mu \nu} (\vp,g) |\Psi\rangle \,,
\label{0.002}
\en
where the source term is given by the expectation value of the energy-momentum tensor operator. 

This semi-classical approximation is non-linear in $\Psi$. However, it is expected that the full quantum theory for gravity will be linear. Consider for example a macroscopic superposition of $\Psi_1$ and $\Psi_2$, where $\Psi_1$ represents a lump of matter in one location and $\Psi_2$ represents that lump in another location. Then it is expected that the full quantum theory for gravity will describe a superposition of the state $\Psi_1$ with its gravitational field and $\Psi_2$ with its gravitational field. This is not the case for the semi-classical approximation. Namely, for the macroscopic superposition $\Psi = (\Psi_1 + \Psi_2)/{\sqrt 2}$, we have that $\langle \Psi| {\widehat T}_{\mu \nu} |\Psi\rangle \approx \left( \langle \Psi_1| {\widehat T}_{\mu \nu} |\Psi_1\rangle + \langle \Psi_2| {\widehat T}_{\mu \nu} |\Psi_2\rangle \right)/2$, so that the gravitational field is affected by two matter sources, one coming from each term in the superposition. As experimentally shown by Page and Geilker, the semi-classical theory becomes inadequate in such situations \cite{page81,kiefer04}. On the other hand it will form a good approximation when the matter state approximately corresponds to a classical state (i.e., a coherent state).

Of course, as already noted by Page and Geilker, it could be that this problem is not due to the fact that gravity is treated classically, but due to the choice of the version of quantum theory. Namely, Page and Geilker adopted the Many Worlds point of view, according to which the wave function never collapses. However, according to standard quantum theory the wave function is supposed to collapse upon measurement. Which physical processes act as measurements is of course rather vague and this is the source of the measurement problem. But it could be that such collapses explain the outcome of their experiment. If an explanation of this type is sought, one should consider so-called spontaneous collapse theories, where collapses are objective, random processes that do not in a fundamental way depend on the notion of measurement. (See \cite{diez-tjedor12} and \cite{derakhshani14,tilloy17} for actual proposals combining such a spontaneous collapse approach with respectively \eqref{0.002} and its non-relativistic version.)

In this paper, we consider an alternative to standard quantum mechanics, called Bohmian mechanics \cite{bohm93,holland93b,duerr09,duerr12}, and develop a semi-classical approximation which is expected to improve upon the usual approach.

Bohmian mechanics solves the measurement problem by introducing an actual configuration (particle positions in the non-relativistic domain, particle positions or fields in the relativistic domain \cite{struyve11a}) that evolves under the influence of the wave function. Instead of coupling classical gravity to the wave function it is now natural to couple it to the actual matter configuration. For example, in the case of a scalar field there is an actual field $\vp_{B}$ whose time evolution is determined by the wave functional $\Psi$. There is an energy-momentum tensor $T_{\mu \nu}(\vp_{B},g)$ corresponding to this scalar field and this tensor can be introduced as the source term in Einstein's field equations:
\be
G_{\mu \nu} (g) = 8\pi G T_{\mu \nu}(\vp_{B},g) \,.
\label{0.003}
\en
While the resulting theory is still non-linear, it solves the problem with the macroscopic superposition. Namely, even though the matter wave function is in a macroscopic superposition of being at two locations, the Bohmian field $\vp_{B}$ and the energy-momentum tensor $T_{\mu \nu}(\vp_{B},g)$ will correspond to a matter configuration at one of the locations. This means that according to eq.\ \eqref{0.003} the gravitational field will correspond to that of matter localised at that location. 

However, there is an immediate problem with this ansatz, namely that eq.\ \eqref{0.003} is not consistent. The Einstein tensor $G_{\mu \nu}$ is identically conserved, i.e., $\nabla^\mu G_{\mu \nu} \equiv 0$. So the Bohmian energy-momentum tensor $T_{\mu \nu}(\vp_{B},g)$ must be conserved as well. However, the equation of motion for the scalar field does not guarantee this. (Similarly, in the Bohmian approach to non-relativistic systems, the energy is generically not conserved. On the other hand, if gravity is also treated quantum mechanically, in the context of the Wheeler-DeWitt theory, then the total energy-momentum tensor is covariantly conserved \cite{pinto-neto19,duerr20b}.) 

We will explain that the root of the problem seems to be the gauge invariance, which in this case is the invariance under spatial diffeomorphisms (i.e., spatial coordinate transformations). The gauge invariance makes that the gauge invariant content is contained in a combination of the metric and scalar field. Therefore, it seems that one can not just assume the metric to be classical without also assuming the scalar field $\vp_{B}$ to be classical (in which case the energy-momentum tensor is conserved). 

We will see that a similar problem arises when we consider a Bohmian semi-classical approach to scalar electrodynamics, which describes a scalar field interacting with an electromagnetic field. In this case, the wave equation for the scalar field is of the form
\be
\ii \pa_t \Psi(\vp,t) = {\widehat H}(\vp,A) \Psi(\vp,t) \,,
\en
where $A$ is the vector potential. There is also a Bohmian scalar field $\vp_{B}$ and a charge current $ j^\nu(\vp_B,A)$ that could act as the source term in Maxwell's equations
\be
\pa_\mu F^{\mu \nu}(A) = j^\nu(\vp_B,A)  \,,
\en
where $F^{\mu \nu}$ is the electromagnetic field tensor. In this case, we have $\pa_\nu \pa_\mu F^{\mu \nu} \equiv 0$ due to the anti-symmetry of $F^{\mu \nu}$. As such, the charge current must be conserved. However, the Bohmian equation of motion for the scalar field does not imply conservation. Hence, just as in the case of gravity, a consistency problem arises. We will find that this problem can be overcome by eliminating the gauge invariance, either by assuming some gauge or (equivalently) by working with gauge-independent degrees of freedom. In this way, we can straightforwardly derive a semi-classical approximation starting from the {\em full} Bohmian approach to scalar electrodynamics (where all degrees of freedom are treated quantum mechanically). For example, in the Coulomb gauge, the result is that there is an extra current $j^\nu_Q$ which appears in addition to the usual (classical) charge current and which depends on the quantum potential, so that Maxwell's equations read 
\be
\pa_\mu F^{\mu \nu}(A) = j^\nu(\vp_B,A) +  j^\nu_Q(\vp_B,A) \,,
\en
which is consistent since the total current is conserved, because of the equation of motion for the scalar field.

While it is easy to eliminate the gauge invariance in the case of electrodynamics, this is notoriously difficult in the case of general relativity. One can formulate a Bohmian theory for the Wheeler-DeWitt approach to quantum gravity, but the usual formulation does not explicitly eliminate the gauge freedom arising from spatial diffeomorphism invariance. Our expectation is that one could find a semi-classical approximation given such a formulation. At least we find our expectation confirmed in simplified symmetry-reduced models, called mini-superspace models, where this invariance is eliminated. We will illustrate this for the model described by the homogeneous and isotropic Friedman-Lema\^itre-Robertson-Walker metric and a uniform scalar field. 

In this paper, we are merely concerned with the formulation of Bohmian semi-classical approximations. Practical applications will be studied elsewhere. Such applications have already been studied for non-relativistic systems in the context of quantum chemistry \cite{gindensperger00,prezhdo01,gindensperger02a,gindensperger02b,meier04,garaschuk11}. It appears that Bohmian semi-classical approximations yield better or equivalent results compared to the usual semi-classical approximation. (They are better in the sense that they are closer to the exact quantum results.) This provides good hope that also in other contexts, such as quantum gravity, the Bohmian approach also gives better results. Potential applications might be found in inflation theory, where the back-reaction from the quantum fluctuations onto the classical background can be studied, or in black hole physics, to study the back-reaction from the Hawking radiation onto space-time.

Other semi-classical approximations have been proposed, see for example \cite{diosi98,hu03,hall05,elze12}, and in particular \cite{burghardt04,burghardt05} where also Bohmian ideas are used. We will not make a comparison with these proposals here.

The semi-classical approximation is just one practical application of Bohmian mechanics. In recent years, others have been explored (see \cite{oriols19,benseny14} for overviews). For example, one may use Bohmian approximation schemes to solve problems in many-body systems \cite{oriols07,albareda09} or one may use Bohmian trajectories as a calculational tool to simulate wave function evolution \cite{wyatt05}. So even though Bohmian mechanics yields the same predictions as standard quantum theory (insofar the latter are unambiguous), it leads to new practical tools and ideas.

The outline of the paper is as follows. We start with an introduction to Bohmian mechanics in section \ref{bm}. In section \ref{scanrqm}, we present the Bohmian semi-classical approximation to non-relativistic quantum theory and its derivation from the full Bohmian theory. In section \ref{scaqam}, we discuss the Bohmian semi-classical approximation to the quantized Abraham model. This model describes extended rigid particles interacting with an electromagnetic field. This is a gauge theory and we will consider two possible gauge fixings, the Coulomb gauge and the temporal gauge. While the former completely fixes the gauge, the latter does not and hence admits a residual gauge symmetry. No consistency problems of the sort mentioned above arise. The reason seems to be that the in the full Bohmian theory the gauge freedom only acts on the electromagnetic field and not on the charges. So no consistency problems arise for the semi-classical approximation where the electromagnetic field is treated classically. Nevertheless, in the case of the temporal gauge, a separation of the remainig gauge freedom from the gauge-invariant degrees of freedom will be important in order to find a good semi-classical approximation. In section \ref{sqed}, we consider scalar electrodynamics, which describes a scalar field coupled to an electromagnetic field. In this case, the gauge freedom acts on both the scalar and electromagnetic field. This leads to consistency problems if the gauge is not completely fixed. First we will consider two gauges, the Coulomb gauge and the unitary gauge, which completely fix the gauge, and show that no consistency problems arise. Then we will consider the temporal gauge which does not completely fix the gauge. In this case, we encounter problems with consistency (due to non-conservedness of the charge current). The temporal gauge is also interesting because it brings scalar electrodynamics in a form which is very similar to that of canonical quantum gravity (described by the Wheeler-DeWitt equation). We turn to gravity in section \ref{qg}. We do not try to develop a Bohmian semi-classical theory from full quantum gravity, but rather make a natural guess for the Bohmian energy-momentum tensor that couples quantum matter to classical gravity. We consider both matter described by a scalar field and by relativistic point-particles. In both cases, the natural guess for the energy-momentum tensor is not conserved. A more careful analysis should start from full quantum gravity. However, because of the spatial diffeomorphism invariance, which acts on both the matter and gravitational field, such an analysis is not expected to help in the construction of a consistent semi-classical approximation, unless a way can be found to gauge fix or to isolate the gauge-independent degrees of freedom. In the case of Newtonian gravity coupled to non-relativistic particles, we can formulate a consistent Bohmian semi-classical approximation, which can be seen as a Bohmian analogue of the Schr\"odinger-Newton equation. Finally, we also consider mini-superspace models for quantum gravity in section \ref{qg}. The model we consider is homogeneous and isotropic and therefore the spatial diffeomorphism invariance is eliminated. In this case, we can derive a consistent semi-classical approximation. We compare this semi-classical approximation with the usual one (which is the equivalent of \eqref{0.001} and \eqref{0.002} for mini-superspace) and show that the latter gives better results.

\section{Bohmian mechanics}\label{bm} 
Non-relativistic Bohmian mechanics (also called pilot-wave theory or \dbb\ theory) is a theory about point-particles in physical space moving under the influence of the wave function \cite{bohm93,holland93b,duerr09,duerr12}. The equation of motion for the configuration $X=({\bf X}_1,\dots,{\bf X}_n)$ of the particles, called the {\em guidance equation}, is given by{\footnote{Throughout the paper we assume units in which $\hbar=c=1$.}}
\be
{\dot X}(t) = v^\psi(X(t),t) \,,
\label{0.01}
\en
where $v^\psi=({\bf v}^\psi_1, \dots , {\bf v}^\psi_n)$, with
\be
{\bf v}^\psi_k = \frac{1}{m_k} {\textrm{Im}}\left( \frac{\boldsymbol{\nabla}_k \psi}{\psi} \right) =  \frac{1}{m_k} {\boldsymbol \nabla}_k S  
\label{0.02}
\en 
and $\psi = |\psi| \ee^{\ii S}$. The wave function $\psi(x,t)=\psi({\bf x}_1,\dots,{\bf x}_n)$ itself satisfies the non-relativistic Schr\"odinger equation
\be
\ii \pa_t \psi(x,t) = \left( - \sum^n_{k=1} \frac{1}{2m_k} \nabla^2_k + V(x) \right) \psi(x,t) \,.
\label{0.03}
\en

For an ensemble of systems all with the same wave function $\psi$, there is a distinguished distribution given by $|\psi|^2$, which is called the {\em quantum equilibrium distribution}. This distribution is {\em equivariant}. That is, it is preserved by the particles dynamics \eqref{0.01} in the sense that if the particle distribution is given by $|\psi(x,t_0)|^2$ at some time $t_0$, then it is given by $|\psi(x,t)|^2$ at all times $t$. This follows from the fact that any distribution $\rho$ that is transported by the particle motion satisfies the continuity equation
\be
\pa_t \rho + \sum^n_{k=1} {\boldsymbol \nabla}_k \cdot ({\bf v}^\psi_k \rho) = 0 
\label{0.04}
\en
and that $|\psi|^2$ satisfies the same equation, i.e.,
\be
\pa_t |\psi|^2 + \sum^n_{k=1} {\boldsymbol \nabla}_k \cdot ({\bf v}^\psi_k |\psi|^2) = 0 \,,
\label{0.041}
\en
as a consequence of the Schr\"odinger equation. It can be shown that for a typical initial configuration of the universe, the (empirical) particle distribution for an actual ensemble of subsystems within the universe will be given by the quantum equilibrium distribution \cite{duerr92a,duerr09,duerr12}. Therefore for such a configuration Bohmian mechanics reproduces the standard quantum predictions. 

Note that the velocity field is of the form $j^\psi/|\psi|^2$, where $j^\psi=({\bf j}^\psi_1,\dots,{\bf j}^\psi_n)$ with ${\bf j}^\psi_k=  {\textrm{Im}}( \psi^* \boldsymbol{\nabla}_k \psi )/m_k $ is the usual quantum current. In other quantum theories, such as for example quantum field theories, the velocity can be defined in a similar way by dividing the appropriate current by the density. In this way equivariance of the density will be ensured. (See \cite{struyve09a} for a treatment of arbitrary Hamiltonians.)

This theory solves the measurement problem. Notions such as measurement or observer play no fundamental role. Instead measurement can be treated as any other physical process.  

There are two aspects of the theory that are important for deriving the semi-classical approximation. Firstly, Bohmian mechanics allows for an unambiguous analysis of the classical limit. Namely, the classical limit is obtained whenever the particles (or at least the relevant macroscopic variables, such as the center of mass) move classically, i.e., satisfy Newton's equation. By taking the time derivative of \eqref{0.01}, we find that
\be
m_k {\ddot {\bf X}}_k(t) = -{\boldsymbol{\nabla}}_k (V(x)+Q^\psi(x,t))\big|_{x=X(t)}\,,
\label{0.07}
\en
where 
\be
Q^\psi = -\sum^n_{k=1}\frac{1}{2m_k}\frac{\nabla^2_k |\psi|}{|\psi|}
\label{0.08}
\en
is the quantum potential. Hence, if the quantum force $-{\boldsymbol{\nabla}}_kQ^\psi$ is negligible compared to the classical force $-{\boldsymbol{\nabla}}_kV$, then the $k$-th particle approximately moves along a classical trajectory. 

Another aspect of the theory is that it allows for a simple and natural definition of the wave function of a subsystem \cite{duerr92a,duerr09}. Namely, consider a system with wave function $\psi(x,y)$ where $x$ is the configuration variable of the subsystem and $y$ is the configuration variable of its environment. The actual configuration is $(X,Y)$, where $X$ is the configuration of the subsystem and $Y$ is the configuration of the other particles. The wave function of the subsystem $\chi(x,t)$, called the {\em conditional wave function}, is then defined as
\be
\chi(x,t) = \psi(x,Y(t),t).
\label{0.05}
\en
This is a natural definition since the trajectory $X(t)$ of the subsystem satisfies 
\be
{\dot X}(t) = v^\psi(X(t),Y(t),t) = v^\chi(X(t),t) \,. 
\label{0.06}
\en
That is, for the evolution of the subsystem's configuration we can either consider the conditional wave function or the total wave function (keeping the initial positions fixed). (The conditional wave function is also the wave function that would be found by a natural operationalist method for defining the wave function of a quantum mechanical subsystem \cite{norsen14}.) The time evolution of the conditional wave function is completely determined by the time evolution of $\psi$ and that of $Y$. The conditional wave function does not necessarily satisfy a Schr\"odinger equation, although in many cases it does. This wave function collapses according to the usual text book rules when an actual measurement is performed.

We will also consider semi-classical approximations to quantum field theories. More specifically, we will consider bosonic quantum field theories. In Bohmian theories it is most easy to introduce actual field variables rather than particle positions \cite{struyve10,struyve11a}. To illustrate how this works, let us consider the free massless real scalar field (for the treatment of other bosonic field theories see \cite{struyve10}). Working in the functional Schr\"odinger picture, the quantum state vector is a wave functional $\Psi(\vp)$ defined on a space of scalar fieldsn $\vp({\bf x})$ in 3-space and it satisfies the functional Schr\"odinger equation
\be
\ii \pa_t \Psi(\vp,t) = \frac{1}{2}\int d^3 x    \left(- \frac{\delta^2}{\delta \vp({\bf x})^2} + {\boldsymbol \nabla} \vp({\bf x}) \cdot{\boldsymbol \nabla} \vp({\bf x}) \right) \Psi(\vp,t) \,.
\label{0.09}
\en
The associated continuity equation is
\be
\pa_t |\Psi(\vp,t)|^2 + \int d^3 x \frac{\delta}{\delta \vp({\bf x})} \left( \frac{\delta S (\vp,t)}{\delta \vp({\bf x})} |\Psi(\vp,t)|^2 \right) = 0 \,,
\label{0.10}
\en
where $\Psi = |\Psi| \ee^{\ii S}$. This suggests the following guidance equation for an actual scalar field $\phi ({\bf x},t)$:
\be
\dot \phi ({\bf x},t) = \frac{\delta S(\vp,t) }{\delta \vp({\bf x})} \bigg|_{\vp({\bf x}) = \phi({\bf x}, t)} \,.
\label{0.11}
\en
Taking the time derivative of this equation results in 
\be
\square \phi({\bf x},t) = - \frac{\delta Q^\Psi(\vp,t)}{\delta \vp({\bf x})}\bigg|_{\vp({\bf x}) = \phi({\bf x}, t)} \,,
\label{0.12}
\en
where
\be
Q^\Psi = - \frac{1}{2|\Psi|} \int d^3 x \frac{\delta^2 |\Psi|}{\delta \vp({\bf x})^2}
\label{0.13}
\en
is the quantum potential. The classical limit is obtained whenever the quantum force, i.e., the right-hand side of eq.\ \eqref{0.12}, is negligible. Then the field approximately satisfies the classical field equation $\square \phi=0$. 

One can also consider the conditional wave functional of a subsystem. A subsystem can in this case be regarded as a system confined to a certain region in space. The conditional wave functional for the field confined to that region is then obtained from the total wave functional by conditioning over the actual field value on the complement of that region. However, in this paper we will not consider this kind of conditional wave functional. Rather, there will be other degrees of freedom, like for example other fields, which will be conditioned over.

This Bohmian theory is not Lorentz invariant. The guidance equation \eqref{0.11} is formulated with respect to a preferred reference frame and as such violates Lorentz invariance. This violation does not show up in the statistical predictions given quantum equilibrium, since the theory makes the same predictions as standard quantum theory which are Lorentz invariant.{\footnote{Actually, this statement needs some qualifications since regulators need to be introduced to make the theory and its statistical predictions well defined \cite{struyve10}.}} The difficulty in finding a Lorentz invariant theory resides in the fact that any adequate formulation of quantum theory must be non-local \cite{goldstein11b}. One approach to make the Bohmian theory Lorentz invariant is by introducing a foliation which is determined by the wave function in a covariant way \cite{duerr14}. In this paper, we will not attempt to maintain Lorentz invariance. As such, the Bohmian semi-classical approximations will not be Lorentz invariant, (very likely) not even concerning the statistical predictions. This is in contrast with the usual semi-classical approach like the one for gravity given by \eqref{0.001} and \eqref{0.002} which is Lorentz invariant. However, this does not take away the expectation that the Bohmian semi-classical approximation will give better or at least equivalent results compared to the usual approach.

\section{Non-relativistic quantum mechanics}\label{scanrqm}

\subsection{Usual versus Bohmian semi-classical approximation}
Consider a composite system of just two particles. The usual semi-classical approach goes as follows. Particle 1 is described quantum mechanically, by a wave function $\chi({\bf x}_1,t)$, which satisfies the Schr\"odinger equation
\be
\ii \pa_t \chi({\bf x}_1,t) =  \left[ - \frac{1}{2m_1}\nabla^2_1  + V({\bf x}_1,{\bf X}_2(t)) \right] \chi({\bf x}_1,t) \,,
\label{1}
\en
where the potential is evaluated at the position of the second particle ${\bf X}_2$, which satisfies Newton's equation
\begin{align}
m_2 {\ddot {\bf X}}_2(t) &=   -\left\langle \chi \left| {\boldsymbol \nabla}_2   V({\bf x}_1,{\bf x}_2)  \big|_{{\bf x}_2={\bf X}_2(t)} \right| \chi \right\rangle \nonumber\\
&= \int d^3x_1|\chi({\bf x}_1,t)|^2  [-{\boldsymbol \nabla}_2 V({\bf x}_1,{\bf x}_2)] \Big|_{{\bf x}_2={\bf X}_2(t)}\,.
\label{2}
\end{align}
So the force on the right-hand-side is averaged over the quantum particle.

An alternative semi-classical approach based on Bohmian mechanics was proposed independently by Gindensperger {\em et al.}\ \cite{gindensperger00} and Prezhdo and Brooksby \cite{prezhdo01}. In this approach there is also an actual position for particle 1, denoted by ${\bf X}_1$, which satisfies the equation
\be
{\dot {\bf X}}_1(t) = {\bf v}^\chi({\bf X}_1(t),t) \,,
\label{3}
\en
where
\be
{\bf v}^\chi = \frac{1}{m_1} {\textrm{Im}} \frac{\bn \chi}{\chi} \,,
\en
and where $\chi$ satisfies the Schr\"odinger equation \eqref{1}. But instead of eq.\ \eqref{2}, the second particle now satisfies 
\be
m_2 {\ddot {\bf X}}_2(t) =   -   {\boldsymbol \nabla}_2   V({\bf X}_1(t),{\bf x}_2)  \big|_{{\bf x}_2={\bf X}_2(t)} \,,
\label{4}
\en
where the force depends on the position of the first particle. So in this approximation the second particle is not acted upon by some average force, but rather by the actual particle of the quantum system. This approximation is therefore expected to yield a better approach than the usual approach, in the sense that it yields predictions closer to those predicted by full quantum theory, especially in the case where the wave function evolves into a superposition of non-overlapping packets. This is indeed confirmed by a number of studies, as we will discuss below.

Let us first mention some properties of this approximation and compare them to the usual approach. In the usual field approach, the specification of an initial wave function $\chi({\bf x},t_0)$, an initial position ${\bf X}_2(t_0)$ and velocity ${\dot {\bf X}}_2(t_0)$ determines a unique solution for the wave function and the trajectory of the classical particle. In the Bohmian approach also the initial position ${\bf X}_1(t_0)$ of the particle of the quantum system needs to be specified in order to uniquely determine a solution. Different initial positions ${\bf X}_1(t_0)$ yield different evolutions for the wave function and the classical particle. This is because the evolution of each of the variables ${\bf X}_1,{\bf X}_2,\chi$ depends on the others. Namely, the evolution of $\chi$ depends on ${\bf X}_2$ via \eqref{1}, whose evolution in turn depends on ${\bf X}_1$ via \eqref{4}, whose evolution in turn depends on $\chi$ via \eqref{3}. (This should be contrasted with the full Bohmian theory, where the wave function acts on the particles, but there is no back-reaction from the particles onto the wave function.) 

The initial configuration ${\bf X}_1(t_0)$ should be considered random with distribution $|\chi({\bf x},t_0)|^2$. However, this does not imply that ${\bf X}_1(t)$ is random with distribution $|\chi({\bf x},t)|^2$ for later times $t$. It is not even clear what the latter statement should mean, since different initial positions ${\bf X}_1(t_0)$ lead to different wave function evolution; so which wave function should $\chi({\bf x},t)$ be? 

This semi-classical approximation has been applied to a number of systems. Prezhdo and Brooksby studied the case of a light particle scattering off a heavy particle \cite{prezhdo01}. They considered the scattering probability over time and found that the Bohmian semi-classical approximation was in better agreement with the exact quantum mechanical prediction than the usual approximation. The Bohmian semi-classical approximation gives probability one for the scattering to have happened after some time, in agreement with the exact result, whereas the probability predicted by the usual approach does not reach one. The reported reason for the better results is that the wave function of the quantum particle evolves into a superposition of non-overlapping packets, which yields bad results for the usual approach (since the force on the classical particle contains contributions from both packets), but not for the Bohmian approach. These results were confirmed and further expanded by Gindensperger {\em et al.}\ \cite{gindensperger02a}. Other examples have been considered in \cite{gindensperger00,gindensperger02b,meier04}. In those cases, the Bohmian semi-classical approximation gave very good agreement with the exact quantum or experimental results. It was always either better or comparable to the usual approach. These results give good hope that the Bohmian semi-classical approximation will also give better results than the usual approximation in other domains such as quantum gravity.

\subsection{Derivation of the Bohmian semi-classical approximation}
The Bohmian semi-classical approach can easily be derived from the full Bohmian theory.{\footnote{The derivation is very close to the one followed by Gindensperger {\em et al.}\ \cite{gindensperger00}. A difference is that they also let the wave function of the quantum system depend parametrically on the position of the classical particle. This leads to a quantum force term in the eq.\ \eqref{4} for particle 2. However, this does not seem to lead to a useful set of equations. In particular, they can not be numerically integrated by simply specifying the initial wave function and particle positions. In any case, Gindensperger {\em et al.}\ drop this quantum force when considering examples \cite{gindensperger00,gindensperger02a,gindensperger02b}, so that the resulting equations correspond to the ones presented above.}} Consider a system of two particles. In the Bohmian description of this system, we have a wave function $\psi({\bf x}_1,{\bf x}_2,t)$ and positions ${\bf X}_1(t),{\bf X}_2(t)$, which respectively satisfy the Schr\"odinger equation
\be
\ii \pa_t \psi = \left[ -\frac{1}{2m_1} \nabla^2_1 - \frac{1}{2m_2} \nabla^2_2 + V({\bf x}_1,{\bf x}_2) \right] \psi 
\label{5}
\en
and the guidance equations
\be
{\dot {\bf X}}_1(t) = {\bf v}^\psi_1({\bf X}_1(t),{\bf X}_2(t),t) \,, \qquad {\dot {\bf X}}_2(t) = {\bf v}^\psi_2({\bf X}_1(t),{\bf X}_2(t),t)\,.
\label{6}
\en
The conditional wave function $\chi({\bf x}_1,t) = \psi({\bf x}_1,{\bf X}_2(t),t)$ for particle 1 satisfies the equation
\be
\ii \pa_t \chi({\bf x}_1,t) =  \left( - \frac{\nabla^2_1}{2m_1}  + V({\bf x}_1,{\bf X}_2(t)) \right) \chi({\bf x}_1,t) + I({\bf x}_1,t) \,,
\label{7}
\en
where
\be
I({\bf x}_1,t) =  \left(- \frac{\nabla^2_2}{2m_2}  \psi({\bf x}_1,{\bf x}_2,t) \right) \Bigg|_{{\bf x}_2={\bf X}_2(t)} +  \ii {\boldsymbol \nabla}_2 \psi({\bf x}_1,{\bf x}_2,t)\Big|_{{\bf x}_2={\bf X}_2(t)} \cdot {\bf v}^\psi_2({\bf X}_1(t),{\bf X}_2(t),t) \,.
\label{8}
\en
So if $I$ is negligible in \eqref{7}, up to a time-dependent factor times $\chi$,{\footnote{If $I$ contains a term of the form $f(t)\chi$, then it can be eliminated by changing the phase of $\chi$ by a time-dependent term.\label{timedependent}}} we are led to the Schr\"odinger equation \eqref{1}. This will for example be the case if $m_2$ is much larger than $m_1$ ($I$ is inversely proportional to $m_2$) and if the wave function slowly varies as a function of ${\bf x}_2$. We also have that
\be
m_2 {\ddot {\bf X}}_2(t) =   - {\boldsymbol \nabla}_2   \left[ V({\bf X}_1(t),{\bf x}_2) + Q^\psi({\bf X}_1(t),{\bf x}_2,t) \right]  \Bigg|_{{\bf x}_2={\bf X}_2(t)}\,,
\en
with $Q^\psi$ the quantum potential. We obtain the classical equation \eqref{4} if the quantum force is negligible compared to the classical force. 

In this way we obtain the equations for a semi-classical formulation. In addition, we also have the conditions under which they will be valid. For other quantum theories, such as quantum gravity, we can follow a similar path to find a Bohmian semi-classical approximation.

\section{The Abraham model}\label{scaqam}
The next theory we consider is the Abraham model, which on the classical level describes extended, rigid, charged particles, which interact with an electromagnetic field \cite{spohn04}. We will consider the Bohmian theory in two gauges: the Coulomb gauge and the temporal gauge. These Bohmian theories are equivalent. An important difference between the gauges is that the former gauge completly fixes the gauge invariance, while the second does not. From the full Bohmian theories, we will derive semi-classical approximations, where the charges are treated quantum mechanically and the electromagnetic field classically. In both cases no problems with consistency arise (like those mentioned in the introduction). The apparent reason is that the gauge transformations only act on the electromagnetic field and not on the charges. Nevertheless, we will see that in the case of the temporal gauge a further separation of gauge degrees of freedom from gauge-invariant degrees of freedom will be desired in order to have a better semi-classical approximation. In section \ref{sqed}, we will consider scalar electrodynamics in those gauges. In that case, because the gauge transformations act on both matter and electromagnetic field, a complete gauge fixing will be necessary to obtain a consistent semi-classical approximation. We end this section with a comparison of the semi-classical theory with similar ones, including one by Kiessling \cite{kiessling07}, which are not derived from the full Bohmian theories.

\subsection{Classical theory}
In the classical Abraham model, the charge distribution of the $i$-th particle is centered around a position ${\bf Q}_i$ and is given by $ eb({\bf x} - {\bf Q}_i)$, where the function $b$ is assumed to be smooth, radial, with compact support and normalized to 1. The particles move under the Lorentz force law
\be
m \ddot {\bf Q}_i(t) = e\left[ {\bf E}_b({\bf Q}_i(t),t)  + \dot {\bf Q}_i(t) \times {\bf B}_b({\bf Q}_i(t),t) \right] \,,
\label{am.01}
\en
where ${\bf E}({\bf x},t)$ and ${\bf B}({\bf x},t)$ are respectively the electric and magnetic field and the subscript $b$ denotes the convolutions
\begin{align}
{\bf E}_b({\bf x},t) &= ({\bf E} * b)({\bf x},t)= \int d^3 y {\bf E}({\bf y},t) b({\bf x} - {\bf y}), \\
{\bf B}_b({\bf x},t) &= ({\bf B} * b)({\bf x},t)= \int d^3 y {\bf B}({\bf y},t) b({\bf x} - {\bf y}). 
\end{align}
In terms of the electromagnetic potential $A^\mu=(A_0,{\bf A})$, the electric and magnetic field are given by ${\bf E}=-\pa_t {\bf A} - \bn A_0$ and ${\bf B} = \bn \times {\bf A}$. The electromagnetic field satisfies Maxwell's equations
\be
\pa_\mu F^{\mu \nu}({\bf x},t) = j^\nu({\bf x},t)\,,
\label{am.02}
\en
with $F^{\mu \nu}=\pa^\mu A^\nu - \pa^\nu A^\mu$ and charge current $j^\nu$ given by 
\be
j^0({\bf x},t) = \sum^n_{i=1} eb({\bf x} - {\bf Q}_i(t)) \,, \qquad {\bf j}({\bf x},t) = \sum^n_{i=1} e \dot {\bf Q}_i(t) b({\bf x} - {\bf Q}_i(t)) \,.
\label{am.03}
\en
Note that although we have employed a covariant notation this model is not covariant, due to the rigidity of the particles. There is a gauge symmetry
\be
A_\mu \to A_\mu - \pa_\mu \al \,.
\label{am.03.01}
\en

We will consider two gauges, namely the Coulomb gauge $\bn \cdot {\bf A}=0$ and the temporal gauge $A_0=0$. For later comparison, we present here the Maxwell's equation in those gauges. In the Coulomb gauge, the equations read
\be
\square {\bf A}^T = {\bf j}^T  \,, \qquad {\bf A}^L=0 \,, \qquad \na^2 A_0 + j^0 = 0 \,, \qquad \bn \dot A_0 = {\bf j}^L \,.
\label{am.03.02}
\en
The superscripts $T$ and $L$ respectively refer to the tranverse and longitudinal part of the vectors.{\footnote{A vector ${\bf V}$ can be written as $ {\bf V}={\bf V}^T + {\bf V}^L$, with ${\bf V}^T={\bf V} - {\boldsymbol{\nabla}} \frac{1}{\nabla^2} {\boldsymbol{\nabla}} \cdot {\bf V}$ and ${\bf V}^L={\boldsymbol{\nabla}} \frac{1}{\nabla^2} {\boldsymbol{\nabla}} \cdot {\bf V}$ respectively the transverse and longitudinal part of the vector potential, where $\frac{1}{\nabla^2}f({\bf x}) = - \frac{1}{4\pi} \int d^3 y \frac{f({\bf y})}{|{\bf x} - {\bf y}|}$. \label{vectorpot}} In the temporal gauge, they read, 
\be
\square {\bf A} + \bn \bn \cdot {\bf A} = {\bf j} \,, \qquad - \bn \cdot {\dot {\bf A}} = j^0 \,.
\label{am.03.03}
\en

\subsection{Coulomb gauge}
\subsubsection{Bohmian theory}
The quantization of the classical model in the Coulomb gauge $\bn \cdot {\bf A}=0$ is straightforward \cite{spohn04}. The wave functional $\Psi(q,{\bf \ca}^T)$, where $q=({\bf q}_1,\dots,{\bf q}_n)$ and ${\bf \ca}^T$ is the transverse part of the vector potential,{\footnote{As before, we will distinguish the arguments $q$ and $\ca^T$ of the wave functional from the actual configuration $Q$ and ${\bf A}^T$.}} satisfies the Schr\"odinger equation{\footnote{We used the notation $\frac{\delta}{ \delta {\bf A}^{T}} = \frac{\delta}{ \delta {\bf A}} - {\boldsymbol{\nabla}} \frac{1}{\nabla^2} {\boldsymbol{\nabla}} \cdot\frac{\delta}{ \delta {\bf A}}$ and $\frac{\delta}{ \delta {\bf A}^{L}} = {\boldsymbol{\nabla}} \frac{1}{\nabla^2} {\boldsymbol{\nabla}} \cdot\frac{\delta}{ \delta {\bf A}}$. \label{derivatives}}} 
\be
\ii \pa_t \Psi = \left[  -\frac{1}{2m} \sum^n_{i=1} \left( \frac{\pa}{\pa {\bf q}_i} - \ii e  {\bf {\mathcal \ca}}^T_b({\bf q}_i) \right)^2 + V_c(q)+  \int d^3 x \left(  - \frac{1}{2} \frac{\delta^2}{ \delta {\bf \ca}^{T}({\bf x})^2} +  \frac{1}{2} (\bn \times {\bf \ca}^T({\bf x}))^2 \right) \right]\Psi,
\label{am.04}
\en
where $V_c$ is the Coulomb potential, which, using
\be
\rho({\bf x},q) = \sum^n_{i=1} eb({\bf x} - {\bf q}_i) 
\label{am.041}
\en
is given by:
\be
V_c(q) = -  \frac{1}{2} \int d^3x \rho({\bf x},q) \frac{1}{\nabla^2} \rho({\bf x},q) = \frac{e^2}{8\pi} \sum^n_{i,j=1} \int d^3xd^3y   \frac{b({\bf x}) b({\bf y}) }{|{\bf x} - {\bf y} + {\bf q}_i -{\bf q}_j |} \,.
\en
So for the electromagnetic field we have used the functional Schr\"odinger representation (see e.g.\ \cite{struyve10} for more details).

In the Bohmian theory the configuration $Q$ and the field ${\bf A}^T$ satisfy 
\be
\dot {\bf Q}_i(t) = \frac{1}{m} \left[ \frac{\pa S(q,{\ca}^T,t)}{\pa {\bf q}_i} - e  \ca^T_b({\bf q}_i) \right]\Bigg|_{Q(t),{\bf A}^T({\bf x},t)} \,, 
\label{am.05}
\en 
\be
\dot {\bf A}^T({\bf x},t) =  \frac{\delta S(q,{\ca}^T,t)}{\delta \ca^T({\bf x})} \Bigg|_{Q(t),{\bf A}^T({\bf x},t)}  \,,
\label{am.06}
\en  
where $\Psi = |\Psi|\ee^{\ii S}$. Taking the time derivative, we find
\be
m \ddot {\bf q}_i(t) = e\left[ {\bf E}_b({\bf q}_i(t),t)  + \dot {\bf q}_i(t) \times {\bf B}_b({\bf q}_i(t),t) \right] - \frac{\pa Q(q,{\bf \ca}^T,t)}{\pa {\bf q}_i} \Bigg|_{Q(t),{\bf A}^T({\bf x},t)} \,,
\label{am.07}
\en
\be
\pa_\mu F^{\mu \nu}({\bf x},t) = j^\nu({\bf x},t) + j^\nu_Q({\bf x},t) \,,
\label{am.08}
\en
where $j^\mu$ is the classical expression for the current given in \eqref{am.03} and 
\be
Q^\Psi = - \frac{1}{2m|\Psi|} \sum^n_{i=1} \frac{\pa^2|\Psi|}{ \pa {\bf q}^2_i} - \frac{1}{2|\Psi|} \int d^3 x  \frac{\delta^2 |\Psi|}{\delta \ca^{T}({\bf x})^2} \,, \quad  j^\mu_Q({\bf x},t) = \left(0,- \frac{\delta Q^\Psi}{\delta \ca^T({\bf x})}\bigg|_{q(t),{\bf A}^T({\bf x},t)} \right)
\label{am.09}
\en
are respectively the quantum potential and what can be called a {\em quantum charge current}, which enters Maxwell's equation in addition to the usual current. Both currents are conserved, i.e., $\pa_\mu j^\mu  = \pa_\mu j^\mu_Q  = 0$. These equations are written in terms of $A^\mu$ and it is assumed that the Coulomb gauge ${\boldsymbol \nabla} \cdot {\bf A}=0$ holds. So the equations \eqref{am.08} are equivalent to
\be
\square {\bf A}^T = {\bf j}^T + {\bf j}_Q \,, \qquad {\bf A}^L=0 \,, \qquad \na^2 A_0 + j^0 = 0 \,, \qquad \bn \dot A_0 = {\bf j}^L \,.
\label{am.11}
\en
The first equation follows from differentiating \eqref{am.06} with respect to time, the second one is implied by the Coulomb gauge, and the third (and the fourth) defines $A_0$ in terms of the charge density. Compared to the classical equations \eqref{am.03.02}, only the first equation has changed through the addition of the quantum current.

\subsubsection{Semi-classical approximation}\label{asca}
We can consider a semi-classical approximation where either the charges or the electromagnetic field behave approximately classically. Let us consider the latter case. We assume the quantum current $j^\nu_Q$ to be negligible in \eqref{am.08}. The conditional wave function for the charges $\chi(q,t)=\psi(q,{\bf A}^T({\bf x},t))$, where $(Q(t),{\bf A}^T({\bf x},t))$ is a particular solution to the guidance equations, then satisfies the equation
\be
\ii \pa_t \chi =  \left[  -\frac{1}{2m} \sum^n_{i=1} \left( \frac{\pa}{\pa {\bf q}_i} - \ii e  {\bf A}^T_b({\bf q}_i,t) \right)^2 + V_c(q) \right] \chi + I \,,
\en
where
\be
I = \int d^3 x \left(  - \frac{1}{2} \frac{\delta^2\Psi}{ \delta \ca^{T}({\bf x})^2} +  \frac{1}{2} (\bn \times \ca^T({\bf x}))^2\Psi   + \ii\frac{\delta \Psi}{ \delta \ca^{T}({\bf x})} \cdot  \dot {\bf A}^T({\bf x},t) \right)\Bigg|_{\ca^T({\bf x})= {\bf A}^T({\bf x},t)} \,.
\en
Whenever $I$ is negligible, up to a time-dependent factor times $\chi$ (cf.\ footnote \ref{timedependent}), the wave equation reduces to 
\be
\ii \pa_t \chi =  \left[  -\frac{1}{2m} \sum^n_{i=1} \left( \frac{\pa}{\pa {\bf q}_i} - \ii e  {\bf A}^T_b({\bf q}_i,t) \right)^2 + V_c(q) \right] \chi \,.
\label{am.15}
\en
Together with
\be
\dot {\bf Q}_i(t) = \frac{1}{m} \left( \frac{\pa S(q,t)}{\pa {\bf q}_i} \Bigg|_{Q(t)} - e  {\bf A}^T_b({\bf Q}_i(t),t) \right) \,, \qquad \pa_\mu F^{\mu \nu}({\bf x},t) = j^\nu({\bf x},t)
\en
and the Coulomb gauge, this defines a consistent semi-classical approximation. That is, due to the anti-symmetry of $F^{\mu \nu}$, we have $\pa_\mu \pa_\nu F^{\mu \nu}=0$ so that the current $j^\nu$ must be conserved, which is actually the case by definition (irrespective of the dynamics of the charges). 

Actually, any current of the form \eqref{am.03} is conserved, irrespective of the dynamics of the charges. The second-order equation for the charges is still of the form \eqref{am.07}, but now with quantum potential
\be
Q^\chi = - \frac{1}{2m|\chi|} \sum^n_{i=1}\frac{\pa^2|\chi|}{ \pa {\bf q}^2_i}  \,.
\en

\subsection{Temporal gauge}
\subsubsection{Bohmian theory}\label{abtemporalbohmian}
The classical theory can also be quantized using the temporal gauge, which is given by $A_0=0$. This gauge does not completely fix the gauge. It still allows for gauge transformations
\be
{\bf A} \to {\bf A} + {\boldsymbol \nabla} \theta \,,
\label{am.15.1}
\en
with $\theta$ time-independent. Quantization yields the Schr\"odinger equation{\footnote{This follows from the usual canonical quantization techniques explained for example in \cite{sundermeyer82}.}}
\be
\ii \pa_t \Psi = \left[  -\frac{1}{2m} \sum^n_{i=1} \left( \frac{\pa}{\pa {\bf q}_i} - \ii e  \ca_b({\bf q}_i) \right)^2 +  \int d^3 x \left(  - \frac{1}{2} \frac{\delta^2}{ \delta \ca({\bf x})^2} +  \frac{1}{2} (\bn \times \ca({\bf x}))^2 \right) \right]\Psi
\label{am.16}
\en
for the wave function $\Psi(q,\ca)$, together with the constraint (corresponding to Gauss's law)
\be
\ii {\boldsymbol \nabla} \cdot \frac{\delta \Psi}{\delta \ca({\bf x})} - \rho({\bf x},q)\Psi =0 \,,
\label{am.17}
\en
with $\rho$ as defined in \eqref{am.041}. The constraint is preserved by the Schr\"odinger dynamics. That is, if it is satisfied at some time, then it is satisfied at all times. The constraint implies the gauge invariance
\be
\Psi(q,\ca) = \ee^{- \ii e\sum^n_{i=1} \theta_b ({\bf q}_i )} \Psi(q,\ca + {\boldsymbol \nabla} \theta)
\label{am.18}
\en
for time-independent $\theta$. 

The guidance equations are
\be
\dot {\bf Q}_i(t) = \frac{1}{m} \left[ \frac{\pa S(q,\ca,t)}{\pa {\bf q}_i}- e  \ca_b({\bf q}_i) \right]\Bigg|_{Q(t),{\bf A}({\bf x},t)} \,, 
\label{am.19}
\en 
\be
\dot {\bf A}({\bf x},t) =  \frac{\delta S(q,\ca,t)}{\delta \ca({\bf x})} \Bigg|_{Q(t),{\bf A}({\bf x},t)}  \,.
\label{am.20}
\en  
Taking the time derivative, we find
\be
m \ddot {\bf Q}_i(t) = e\left[ {\bf E}_b({\bf Q}_i(t),t)  + \dot {\bf Q}_i(t) \times {\bf B}_b({\bf Q}_i(t),t) \right] - \frac{\pa Q^\Psi(q,\ca^T,t)}{\pa {\bf q}_i}\Bigg|_{Q(t),{\bf A}^T({\bf x},t)} \,,
\label{am.21}
\en
\be
\pa_\mu F^{\mu \nu} = j^\nu + j^\nu_Q \,,
\label{am.22}
\en
where $A^\mu=(0,{\bf A})$, $j^\mu$ is the classical expression for the current given in \eqref{am.03}, and 
\be
Q^\Psi = - \frac{1}{2m|\Psi|} \sum^n_{i=1} \frac{\pa^2|\Psi|}{\pa {\bf q}^2_i} - \frac{1}{2|\Psi|} \int d^3 x  \frac{\delta^2 |\Psi|}{\delta \ca({\bf x})^2} \,, \quad  j^\mu_Q({\bf x},t) = \left(0,- \frac{\delta Q^\Psi}{\delta \ca({\bf x})}\bigg|_{Q(t),{\bf A}({\bf x},t)} \right) \,.
\label{am.23}
\en
More in detail, \eqref{am.22} can be written as
\be
\square {\bf A} + \bn \bn \cdot {\bf A} = {\bf j}+ {\bf j}_Q \,, \qquad - \bn \cdot {\dot {\bf A}} = j^0 \,,
\label{am.24}
\en
where the first equation follows from taking the time derivative of \eqref{am.20} and the second equation follows from \eqref{am.20} and the constraint \eqref{am.17}. The difference with the classical equations \eqref{am.03.03} is the presence of the quantum current ${\bf j}_Q$. 

Because of the identity \eqref{am.18}, the Bohmian theory is invariant under the time-independent gauge transformations \eqref{am.15.1}.

This Bohmian theory is equivalent to the one formulated in the Coulomb gauge. To see this, write $\ca = \ca^T + \ca^L$ and define 
\be
\Psi'(q,\ca^T) = \ee^{- \ii \int d^3 x\bn \cdot \ca({\bf x}) \frac{1}{\nabla^2} \rho({\bf x},q)} \Psi(q,\ca + {\boldsymbol \nabla} \theta)= \ee^{- \ii e \sum^n_{i=1}  \frac{1}{\nabla^2} \bn \cdot \ca_b({\bf q}_i) } \Psi(q,\ca + {\boldsymbol \nabla} \theta) \,.
\en 
The constraint \eqref{am.17} then implies that $\delta \Psi' / \delta \ca^L = 0$ so that $\Psi'$ is indeed independent of $\ca^L$. It is easy to verify that $\Psi'$ satisfies the Schr\"odinger equation in the Coulomb gauge. The appearance of the Coulomb potential $V_c(q)$ follows from the constraint \eqref{am.17} which implies that
\be
- \frac{1}{2}\int d^3 x \frac{\delta^2 \Psi}{ \delta \ca^{L}({\bf x})^2} =  - \frac{1}{2} \int d^3 x \rho({\bf x},q) \frac{1}{\nabla^2}\rho({\bf x},q)\Psi  = V_c(q)\Psi\,.
\label{am.30}
\en
The guidance equations \eqref{am.19} and \eqref{am.20} respectively reduce to \eqref{am.05} and \eqref{am.06}. The guidance equation \eqref{am.20} also yields $-{\boldsymbol \nabla} \cdot \dot{\bf A}^L= j^0$. This means that ${\bf A}^L$ is determined by $j^0$ up to time-independent gauge transformation. As such, ${\bf A}^L$ is not an independent dynamical degree of freedom and could be ignored. Also after quantization in the Coulomb gauge, there is no degree of freedom correpsonding to ${\bf A}^L$ and hence we can conclude that both gauges lead to equivalent Bohmian theories.

\subsubsection{Semi-classical approximation}\label{abrahamtempsca}
Again, the derivation of the semi-classical approximation can be carried out by considering the conditional wave function $\chi(q,t)=\psi(q,{\bf A}({\bf x},t))$. There is one term in $\ii \pa_t \chi$ that requires special attention, namely
\be
I = \int d^3 x \left(  - \frac{1}{2} \frac{\delta^2\Psi}{ \delta \ca({\bf x})^2}    + \ii\frac{\delta \Psi}{ \delta \ca({\bf x})} \cdot  \dot {\bf A}({\bf x},t) \right)\Bigg|_{\ca({\bf x})= {\bf A}({\bf x},t)} \,.
\en
One might be tempted to ignore this term. However, because of the constraint \eqref{am.17}, further analysis is needed. Writing the vector potential in terms of its tranverse and longitudinal component and using the identities \eqref{am.30} and
\be
\int d^3 x \ii \frac{\delta \Psi}{ \delta \ca^L({\bf x})} \Bigg|_{\ca({\bf x})= {\bf A}({\bf x},t)}\cdot  \dot {\bf A}^L({\bf x},t)  = \int d^3 x \rho({\bf x},q) \frac{1}{\nabla^2} j^0({\bf x},t) \chi \,,
\en
which both follow from the constraint, we get that 
\begin{multline}
I = \int d^3 x \left(  - \frac{1}{2} \frac{\delta^2\Psi}{ \delta \ca^{T}({\bf x})^2} + \ii\frac{\delta \Psi}{ \delta \ca^{T}({\bf x})} \cdot  \dot {\bf A}^{T}({\bf x},t) \right)\Bigg|_{\ca^T({\bf x})= {\bf A}^T({\bf x},t)} \\
+ \int d^3 x  \rho({\bf x},q) \frac{1}{\nabla^2} \left( j^0({\bf x},t) - \frac{1}{2} \rho({\bf x},q) \right) \chi \,.
\end{multline}
While the first term should be ignored in the semi-classical approximation, the second term is best kept. As a result the wave equation reads
\be
\ii \pa_t \chi =    -\frac{1}{2m} \sum^n_{i=1} \left( \frac{\pa}{\pa {\bf q}_i} - \ii e  {\bf A}_b({\bf q}_i,t) \right)^2  \chi+ \int d^3 x  \rho({\bf x},q) \frac{1}{\nabla^2} \left( j^0({\bf x},t) - \frac{1}{2} \rho({\bf x},q) \right) \chi \,.
\label{am.40}
\en
Since $j^0({\bf x},t) = \rho({\bf x},Q(t)) $, the second term may presumably be approximated by $-V_c(q)$ for practical purposes. 

The semi-classical theory is completed by the following equations for the configuration $Q$ and the potential $A^\mu=(0,{\bf A})$: 
\be
\dot {\bf Q}_i(t) = \frac{1}{m} \left( \frac{\pa S(q,t)}{\pa {\bf q}_i} \Bigg|_{Q(t)} - e  {\bf A}_b({\bf Q}_i(t),t) \right) \,, \qquad \pa_\mu F^{\mu \nu}({\bf x},t) = j^\nu({\bf x},t) \,.
\en
This theory is consistent since $j^\nu$ is conserved by definition. It is invariant under the gauge symmetry ${\bf A} \to {\bf A} + {\boldsymbol \nabla} \theta$, $\Psi(q) \to \ee^{ \ii e\sum^n_{i=1} \theta_b ({\bf q}_i )} \Psi(q)$. It is also equivalent to the semi-classical theory in the Coulomb gauge. This is shown similarly as for the full Bohmian theory, by considering again the decomposotion ${\bf A} = {\bf A}^T + {\bf A}^L$ and by defining
\be
\Psi'(q,t) = \ee^{- \ii \int d^3 x\bn \cdot {\bf A}({\bf x},t) \frac{1}{\nabla^2} \rho({\bf x},q)} \Psi(q,t) = \ee^{- \ii e \sum^n_{i=1}  \frac{1}{\nabla^2} \bn \cdot {\bf A}_b({\bf q}_i,t) } \Psi(q,t) \,.
\en
which satisfies \eqref{am.15}.

There are two important observations to make. First, we did not run into any problems of the type mentioned in the introduction concerning the consistency of the equations of motion, even though the full Bohmian theory is a gauge theory, with gauge symmetry given by \eqref{am.15.1}. The reason seems to be that the gauge symmetry acts solely on the electromagnetic field and leaves the matter degrees of freedom invariant, rather than mixing the degrees of freedom. Therefore we can consider a consistent approximation where the electromagnetic field is treated classically (even if we had ignored the potential term in \eqref{am.40}). In the next section, we will consider a matter field instead of particles. In that case we run into problems with consistency if the gauge is not completely fixed or if the gauge-invariant degrees are not separated from the gauge degrees of freedom.

Second, to get the appropriate Schr\"odinger equation we have used the separation into transverse and longitudinal parts of the vector potential. The transverse part is gauge invariant, while the longitudal part is a gauge degree of freedom. So in effect, we have separated the gauge degrees of freedom form the gauge-independent degrees of freedom. While this was not necessary for consistency, it seems necessary to obtain a better semi-classical approximation.

\subsection{Alternative models}\label{ascaam}
We can actually consider different semi-classical models that are consistent. Namely, consider 
\be
\ii \pa_t \chi =  \left[  -\frac{1}{2m} \sum^n_{i=1} \left(\frac{\pa }{\pa {\bf q}_i}  - \ii e  {\bf A}_b({\bf q}_i,t) \right)^2 + V_b(q,t) \right] \chi \,,
\en
\be
\dot {\bf Q}_i(t) = \frac{1}{m} \left( \frac{\pa S(q,t)}{\pa {\bf q}_i} \Bigg|_{Q(t)}  - e  {\bf A}_b({\bf Q}_i(t),t) \right) \,, \qquad \pa_\mu F^{\mu \nu}({\bf x},t) = j^\nu({\bf x},t) \,,
\en
where $V_b(q,t) = \sum^n_{i=1} e A_{0 b}({\bf q}_i,t)$. Kiessling considered such a model assuming the Lorenz gauge $\pa_\mu A^\mu$ \cite{kiessling07}.{\footnote{Kiessling also considers charges with spin and possible alternative guidance equations.}} We could also consider the Coulomb gauge ${\boldsymbol \nabla} \cdot {\bf A} = 0$. However, in the latter case, the semi-classical approximation is different from the one presented in section \ref{asca}. The difference lies in the scalar potential in the Schr\"odinger equation. In the latter case, the potential is given $V_c(q)$, which is a function {\em independent} of the actual positions $Q_i(t)$, whereas in the former case, the potential $V_b(q,t)$ {\em depends} on the actual positions because of Gauss's law $\nabla^2 A_0({\bf x},t) + j^0({\bf x},t)=0$. The difference may however be negligible for practical applications. In the case of the temporal gauge, we do not get the potential term that is in \eqref{am.40}. So here we see the virtue of starting from the full Bohmian theory in order to derive a semi-classical approximation.

\section{Scalar electrodynamics}\label{sqed}
Scalar electrodynamics describes a scalar field interacting with an electromagnetic field. There is a gauge symmetry which transforms both the scalar field and electromagnetic field. There exist various different but equivalent formulations of Bohmian scalar electrodynamics \cite{struyve10}. These formulations can be found either by considering different gauges or by working with different choices of gauge-independent variables (which more or less amounts to the same thing \cite{struyve10,struyve11b}). The gauges we will consider here are the Coulomb gauge, the unitary gauge and the temporal gauge. However, while the Coulomb and the unitary gauge completely fix the gauge freedom, the temporal gauge does not. We will see that semi-classical approximations can straightforwardly be found in the case of the Coulomb and the unitary gauge, while problems with consistency arise in the case of the temporal gauge. The problem seems to originate from the fact that there is a remaining gauge symmetry for the scalar and electromagnetic fields, such that the gauge invariant content lies in a combination of both. So it seems necessary to completely eliminate the gauge freedom (or at least separate gauge degrees of freedom from gauge-independent degrees of freedom). 

\subsection{Coulomb gauge}\label{coulomb} 
\subsubsection{Classical theory}
The Lagrangian fpr scalar electrodynamics is
\begin{equation}
L =  \int d^3 x \left((D_{\mu} \phi)^* D^{\mu} \phi - m^2\phi^*\phi - \frac{1}{4} F^{\mu \nu}F_{\mu \nu} \right)  \,,
\label{sq.1}
\end{equation}
where $D_{\mu}= \partial_{\mu} + \ii e A_{\mu}$ is the covariant derivative, with $A^\mu=(A_0,{\bf A})$ and $F^{\mu \nu} = \pa^\mu A^\nu - \pa^\nu A^\mu$. The corresponding field equations are 
\begin{equation}
D_{\mu}D^{\mu}\phi + m^2 \phi = 0\,, \qquad \partial_{\mu} F^{\mu \nu}= j^{\nu} \,,
\label{sq.2}
\end{equation}
where 
\be
j^{\nu} = \ii e \left(\phi^*D^{\nu}\phi - \phi D^{\nu*} \phi^* \right)
\label{sq.3}
\en
is the charge current. The theory has a local gauge symmetry
\begin{equation}
\phi \to e^{\ii e \alpha} \phi \,,  \quad  A^{\mu} \to   A^\mu -  \partial^\mu  \alpha\,. 
\label{sq.4}
\end{equation}

In this section we will focus on the Coulomb gauge ${\boldsymbol \nabla} \cdot {\bf A}=0$. Writing ${\bf A}={\bf A}^T + {\bf A}^L$ (cf.\ footnote \ref{vectorpot}), this gauge corresponds to ${\bf A}^L = 0$ and the equations of motion reduce to
\be
\left(\square   + 2\ii e A_0 \pa_t + \ii e {\dot A}_0  - e^2 A^2_0  + 2\ii e {\bf A}^T \cdot {\boldsymbol{\nabla}}  + e^2 {\bf A}^{T2} + m^2 \right) \phi = 0\,, 
\label{sq.5}
\en
\be
\square {\bf A}^T + \bn {\dot A}_0 = {\bf j} \,, 
\label{sq.5.1}
\en
\be
-\nabla^2 A_0 =  j_0  \,,
\label{sq.6}
\en
with 
\be
{\bf j} = \ii e (\phi \bn \phi^* - \phi^* \bn \phi) - 2 e^2 {\bf A}^T |\phi|^2 \,, 
\label{sq.6.1}
\en
\be
j_0 = \ii e(\phi^* {\dot \phi} - \phi{\dot \phi}^*) - 2 e^2 A_0 |\phi|^2 \,. 
\label{sq.6.2}
\en
Eq.\ \eqref{sq.6} can be written as 
\be 
\left(\nabla^2  - 2 e^2 |\phi|^2\right) A_0  = - \ii e(\phi^* {\dot \phi} - \phi{\dot \phi}^*) \,.
\label{sq.7}
\en
This is not a dynamical equation but rather a constraint on $A_0$; it could be used to solve for $A_0$ in terms of $\phi$.

\subsubsection{Bohmian theory}
The classical theory can easily be quantized using the Coulomb gauge. The same quantum field theory can be obtained by eliminating the gauge degrees of freedom on the classical level and then quantizing the remaining gauge-independent degrees of freedom \cite{struyve10}. In the resulting Bohmian theory, the actual fields $\phi$ and ${\bf A}^T$ are guided by a wave functional $\Psi(\vp, \ca^T,t)$ which satisfies the functional Schr\"odinger equation{\footnote{The wave functional should be understood as a functional of the real and imaginary part of $\vp$. In addition, writing $\vp = (\vp_r + \ii \vp_i)/{\sqrt 2}$, we have that the functional derivatives are given by Wirtinger derivatives $\delta / \delta \vp = (\delta / \delta \vp_r  - \ii \delta / \delta \vp_i)/{\sqrt 2}$ and $\delta / \delta \vp = (\delta / \delta \vp_r  + \ii \delta / \delta \vp_i)/{\sqrt 2}$.}}
\be
\ii \pa_t \Psi =   \int d^3 x \bigg(-\frac{ \delta^2  }{ \delta \vp^* \delta \vp} + | (\bn - \ii e \ca^T) \vp|^2  + m^2 |\vp|^2     
- \frac{1}{2} {\mathcal C} \frac{1}{\nabla^2} {\mathcal C}  
 - \frac{1}{2} \frac{\delta^2}{ \delta \ca^{T2}} + \frac{1}{2}(\bn \times \ca^T)^2 \bigg)\Psi \,, 
\label{sq.8}
\en
where
\be
{\mathcal C} ({\bf x}) = e \left(\vp^*({\bf x}) \frac{ \delta  }{\delta \vp^*({\bf x})} -  \vp({\bf x}) \frac{ \delta }{ \delta \vp({\bf x})} \right)
\label{sq.8.001}
\en
is the charge density operator in the functional Schr\"odinger picture. The first three terms in the Hamiltonian correspond to the Hamiltonian of a scalar field minimally coupled to a transverse vector potential. The fourth term corresponds to the Coulomb potential and the remaining terms to the Hamiltonian of a free electromagnetic field. 
 
The guidance equations are
\be
{\dot\phi } =  \left( \frac{\delta S}{\delta \vp^* }  - e \vp \frac{1}{\nabla^2}{\mathcal C} S\right)\Bigg|_{\phi,{\bf A}^T} \,,\qquad \dot {\bf A}^T =  \frac{\delta S}{\delta \ca^T  }\Bigg|_{\phi,{\bf A}^T} \,,
\en
with $\Psi = |\Psi|\ee^{\ii S} $. Defining 
\be
A_0 = - \ii \frac{1}{\nabla^2}{\mathcal C} S\Bigg|_{\phi,{\bf A}^T} \,,
\en
we can rewrite the guidance equation for the scalar field as
\be
D_0 \phi = \frac{\delta S}{\delta \vp^*}\Bigg|_{\phi,{\bf A}^T} \,.
\en
With this definition of $A_0$, the classical equation \eqref{sq.7} is satisfied.

Taking the time derivative of the guidance equations we obtain, after some calculation, that
\be
D_\mu D^\mu \phi - m^2 \phi = - \frac{\delta Q^\Psi}{\delta \vp^*}\Bigg|_{\phi,{\bf A}^T} \,,
\label{sq.8.01}
\en
\be
\pa_\mu F^{\mu \nu} = j^\nu + j^\nu_Q\,,
\label{sq.8.1}
\en
where $A^\mu=(A_0,{\bf A}^{T})$. The quantum potential is given by 
\be
Q^\Psi = - \frac{1}{|\Psi|} \int d^3 x \left( \frac{\delta^2 }{\delta \vp^* \delta \vp} +\frac{1}{2} {\mathcal C} \frac{1}{\nabla^2} {\mathcal C}  + \frac{1}{2} \frac{\delta^2 }{\delta \ca^{T2}} \right)|\Psi| \,.
\label{sq.9}
\en
The current $j^\mu$ is given by the classical expression \eqref{sq.6.1} and \eqref{sq.6.2}. $j^\nu_Q = (0,{\bf j}_Q)$, with
\be
{\bf j}_Q= \left( \ii {\boldsymbol \nabla} \frac{1}{\nabla^2}{\mathcal C} Q^\Psi - \frac{\delta Q^\Psi}{\delta\ca^T }\right)\Bigg|_{\phi,{\bf A}^T} \,,
\label{sq.10}
\en
can be considered an additional quantum current. The total current $j^\nu + j^\nu_Q$ is conserved, i.e., $\pa_\nu (j^\nu + j^\nu_Q) = 0$, as a consequence of \eqref{sq.8.01}. This is required for consistency of \eqref{sq.8.1}, since $ \pa_\nu \pa_\mu F^{\mu \nu} \equiv 0$ (due to the anti-symmetry of $F^{\mu \nu}$). In contrast, the current $j^\nu$ satisfies $\pa_\nu j^\nu = -\ii {\mathcal C} Q^\Psi|_{\phi,{\bf A}^T}$ and is hence not necessarily conserved.

\subsubsection{Semi-classical approximation: classical electromagnetic field}\label{cgscacef}
Let us first consider a semi-classical approximation by treating the electromagnetic field classically and the scalar field quantum mechanically. Similarly as before, it can be found by considering the conditional wave functional $\chi(\vp,t)=\Psi(\vp,{\bf A}^T(t),t)$ for the scalar field and conditions under which the electromagnetic field approximately behaves classically. We will not present this procedure here, but just give the resulting equations.

The wave functional $\chi(\vp,t)$ satisfies the functional Schr\"odinger equation for a quantized scalar field moving in an external classical transverse vector potential and Coulomb potential:
\be
\ii \pa_t \chi =   \int d^3 x \bigg(-\frac{ \delta^2  }{ \delta \vp^* \delta \vp} +  | (\bn - \ii e {\bf A}^T) \vp|^2 + m^2 |\vp|^2  - \frac{1}{2} {\mathcal C} \frac{1}{\nabla^2} {\mathcal C} \bigg)\chi \,,   
\label{sq.11}
\en
where ${\mathcal C}$ is defined as before. The actual scalar field satisfies
\be
D_0 \phi = \frac{\delta S}{\delta \vp^*}\Bigg|_{\phi} \,,
\label{sq.12}
\en
where $A_0$ is defined as before, with $S$ now the phase of $\chi$. The vector potential $A^\mu = (A_0, {\bf A}^T)$ satisfies Maxwell's equations
\be
\pa_\mu F^{\mu \nu} = j^\nu + j^\nu_Q\,,
\label{sq.13}
\en
where $j^\nu_Q = (0,{\bf j}_Q)$ is an additional quantum current, with
\be
{\bf j}_Q=  \ii {\boldsymbol \nabla} \frac{1}{\nabla^2}{\mathcal C} Q^\chi \Bigg|_{\phi} 
\en
and 
\be
Q^\chi= - \frac{1}{|\chi|} \int d^3 x \left( \frac{\delta^2}{\delta \vp^* \delta \vp} + \frac{1}{2}{\mathcal C} \frac{1}{\nabla^2} {\mathcal C}  \right)|\chi|\,.
\en
So the equations \eqref{sq.11}-\eqref{sq.13} define the semi-classical approximation. 

We still have that $\pa_\mu (j^\mu + j^\mu_Q) = 0$, so that the Maxwell equations \eqref{sq.13} are consistent. The correct source term in Maxwell's equations was found by considering the full Bohmian theory. Just using the current $j^\mu$ would yield an inconsistent set of equations since $\pa_\nu j^\nu = -\ii {\mathcal C} Q^\chi|_{\phi}$. Adding an extra current to $j^\mu$ so that the total one would be conserved would still leave an ambiguity since one could always add a vector that is conserved. 

Another essential ingredient in our derivation was that the gauge freedom was eliminated. We will see in section \ref{temporal} that without such an elimination it does not seem possible to derive a semi-classical approximation.

\subsubsection{Semi-classical approximation: classical scalar field}
We can also consider a semi-classical approximation where the scalar field is treated classically and the electromagnetic field quantum mechanically. In this case, the functional Schr\"odinger equation for the wave function $\chi(\ca^{T},t)$ is
\be
\ii \pa_t \chi =   \int d^3 x \bigg( - \frac{1}{2} \frac{\delta^2}{ \delta \ca^{T2} }  + \frac{1}{2} (\bn \times \ca^T)^2  +\ii e \ca^T \cdot  \left(\phi^* \bn \phi -\phi \bn  \phi^* \right) + e^2\ca^{T2} | \phi|^2    \bigg)\chi \,.
\label{sc.sf.1}
\en
The actual potential ${\bf A}^T$ satisfies the guidance equation 
\be
\dot {\bf A}^T =  \frac{\delta S}{\delta \ca^T  } \Bigg|_{{\bf A}^T}
\label{sc.sf.2}
\en
and the scalar field satisfies the classical equation
\be
D_\mu D^\mu \phi - m^2 \phi = 0 \,,
\label{sc.sf.5}
\en
where again $A^\mu = (A_0, {\bf A}^T)$ with $A_0$ defined by \eqref{sq.7}.

\subsection{Unitary gauge}\label{unitary} 
There are other possible semi-classical approximations that can be considered. We have 4 independent field degrees of freedom, namely the two transverse degrees of freedom of the vector potential and the two (real) degrees of freedom of the scalar field, and any of these or combinations thereof could in principle be assumed classical. Here, we will consider two different semi-classical approximations. One where the amplitude of the scalar field is assumed classical and another one where all degrees of freedom but the amplitude are assumed classical. While we could consider these approximations using the Bohmian formulation in terms of the Coulomb gauge, it is more elegant to consider it in a different yet equivalent formulation which can be obtained by imposing the unitary gauge. As in the case of the Coulomb gauge no consistency problems arise. 

\subsubsection{Classical theory}
The unitary gauge is given by $\phi = \phi^*$. Writing 
\be
\phi = \ue \ee^{\ii \theta} / {\sqrt 2} \,,   
\label{u.0}
\en
with $\ue= {\sqrt 2} |\phi|$ (which can be done where $\phi \neq 0$), this gauge amounts to $\theta=0$. The classical equations become
\be
\square \ue + m^2 \ue - e^2  A^\mu A_\mu \ue = 0 \,, \quad \pa_\mu F^{\mu \nu} = j^\nu\,,
\label{u.1}
\en
where now $j^\nu = -e^2 \ue^2 A^\nu$. Maxwell's equations become
\be
\square {\bf A} + \bn \bn \cdot {\bf A} + e^2 \ue^2 {\bf A} + \bn {\dot A}_0  = 0 \,, 
\label{u.1.2}
\en
\be
\left( \nabla^2 - e^2 \ue^2 \right) A_0 + \bn \cdot {\dot {\bf A}} = 0 \,.
\label{u.2}
\en
The last equation is again a constraint rather than a dynamical equation; it can be used to express $A_0$ in terms of ${\bf A}$.

\subsubsection{Bohmian theory}
Quantization of the classical theory leads to the following functional Schr\"odinger equation for $\Psi(\eta,\ca)$:
\begin{multline} 
\ii \pa_t \Psi =  \frac{1}{2}  \int d^3 x \bigg( -  \frac{1}{\eta}\frac{ \delta  }{ \delta \eta}\left(\eta  \frac{ \delta  }{ \delta \eta} \right) + (\bn \eta)^2 + m^2 \eta^2 +  e^2  \ca^2 \eta^2\\
-  \frac{\delta^2}{ \delta \ca^2} -\frac{1}{e^2 \eta^2}\left( \bn \cdot \frac{\delta}{  \delta \ca}\right)^2 + (\bn \times \ca)^2 \bigg)\Psi \,.
\label{u.3}
\end{multline} 
The guidance equations are
\be
{\dot \ue} = \frac{\delta S}{\delta \eta }\Bigg|_{\ue,{\bf A}} \,, \qquad \qquad \dot {\bf A} =  \left[\frac{\delta S}{\delta  \ca } - \bn \left(\frac{1}{e^2 \eta^2} \bn \cdot \frac{\delta S}{\delta \ca} \right)\right]\Bigg|_{\ue,{\bf A}} \,.
\label{u.4}
\en
Defining 
\be
A_0 = \frac{1}{e^2 \ue^2} \bn \cdot \frac{\delta S}{\delta \ca}\Bigg|_{\ue,{\bf A}} \,,
\label{u.5}
\en
the latter equation can be written as $\dot {\bf A} =(\delta S / \delta \ca) |_{\ue,{\bf A}}- \bn A_0$. With this definition the classical equation \eqref{u.2} also holds for the Bohmian dynamics.

The second-order equations are
\be
\square \ue + m^2 \ue - e^2  A^\mu A_\mu \ue = - \frac{\delta Q^\Psi}{\delta \eta}\Bigg|_{\ue,{\bf A}} \,, \quad \pa_\mu F^{\mu \nu} =  j^\nu + j^\nu_Q \,,
\en
where $j^\nu_Q = (0,{\bf j}_Q)$, with
\be
{\bf j}_Q  = - \frac{\delta Q^\Psi}{\delta \ca}\Bigg|_{\ue,{\bf A}} \,,
\en
and
\be
Q^\Psi = - \frac{1}{2|\Psi|} \int d^3 x \left( \frac{1}{\eta} \frac{\delta}{ \delta \eta} \left(\eta  \frac{ \delta  }{ \delta \eta} \right)
+ \frac{\delta^2 }{ \delta \ca^2} + \frac{1}{e^2 \eta^2}\left( \bn \cdot \frac{\delta}{  \delta \ca}\right)^2 \right)|\Psi| \,.
\en

This Bohmian theory is equivalent to the one considered in the previous section. This can easily be checked by using the field transformation $(\phi,{\bf A}^T) \to (\ue, {\bf A})$ which is obtained by using the polar decomposition \eqref{u.0} for $\phi$ and ${\bf A} = {\bf A}^T + \frac{1}{e} \bn \theta$ (with inverse transformation $\theta = e \frac{1}{\nabla^2} \bn \cdot {\bf A}$).{\footnote{The equivalence of the Schr\"odinger picture for the Coulomb gauge and the unitary gauge was considered before in \cite{kim90} (but seems to require a small correction for the kinetic term for $\ue$).}}

\subsubsection{Semi-classical approximation: classical electromagnetic field}
When the vector potential ${\bf A}$ evolves approximately classically, we have the following semi-classical approximation. The wave functional $\chi(\eta,t)$ satisfies
\be
\ii \pa_t \chi =  \frac{1}{2}  \int d^3 x \bigg( -  \frac{1}{\eta}\frac{ \delta  }{ \delta \eta}\left(\eta  \frac{ \delta  }{ \delta \eta} \right) +  (\bn \eta)^2 + m^2 \eta^2 -  e^2  A_\mu A^\mu \eta^2  \bigg) \chi 
\label{u.10}
\en
and the guidance equation
\be
{\dot \ue} = \frac{\delta S}{\delta \eta }\Bigg|_{\ue} \,.
\label{u.11}
\en
The electromagnetic field satisfies
\be
\pa_\mu F^{\mu \nu} = -e^2 \ue^2 A^\nu \,.
\label{u.12}
\en

The appearance of the term containing $A_0$ in the Schr\"odinger equation requires some explanation. As before, the equation can be obtained by assuming that certain terms negligible when considering the time derivative of the conditional wave functional $\chi(\eta,t)=\Psi(\eta,{\bf A}(t),t)$. In this case, $\ii \pa_t \chi$ contains the term
\begin{multline}
- \int d^3 x \frac{1}{2e^2 \eta^2({\bf x})}  \left(\bn \cdot \frac{\delta S}{  \delta \ca(\bf x)}\right)^2\Bigg|_{{\bf A}({\bf x},t)} \chi= \\
 \int d^3 x  \frac{e^2 \eta^2({\bf x})}{2}  \left\{ - A^2_0({\bf x},t)  +  \left[ A^2_0({\bf x},t) -\frac{1}{e^4 \eta^4({\bf x})}  \left(\bn \cdot \frac{\delta S}{  \delta \ca(\bf x)}\right)^2\Bigg|_{ {\bf A}({\bf x},t)} \right] \right\}\chi \,.
\end{multline}
The term within the big square brackets on the right hand side would be zero when it is evaluated for the actual field $\ue({\bf x},t)$ because of \eqref{u.5}. Otherwise, it is not necessarily zero. In our semi-classical approximation we assume that this term is negligible. As such, the resulting Schr\"odinger equation \eqref{u.10} corresponds to the one of a quantized scalar field minimally coupled to a classical electromagnetic field. 

Note that there is no consistency issue in this case. Just as in the classical or the full Bohmian case, the equation $\pa_\mu j^\mu= \pa_\mu (-e^2 \ue^2 A^\mu) =0$ does not follow from the equation of $\ue$, but from the Maxwell equations themselves.

\subsubsection{Semi-classical approximation: classical scalar field}
In the case the scalar field behaves approximately classically, we have the following semi-classical approximation. The wave functional $\chi(\ca,t)$ satisfies
\be
\ii \pa_t \chi =    \frac{1}{2} \int d^3 x \bigg( -  \frac{\delta^2}{ \delta \ca^2} -\frac{1}{e^2 \ue^2}\left( \bn \cdot \frac{\delta}{  \delta \ca}\right)^2 +  (\bn \times \ca)^2 + e^2\ue^2  \ca^2  \bigg)\chi 
\en
and guides the field ${\bf A}$ through the guidance equation
\be
\dot {\bf A} =\frac{\delta S}{ \delta \ca}\Bigg|_{{\bf A}} - \bn A_0 \,,
\en
where $A_0$ is defined as in \eqref{u.5}. The field $\ue$ satisfies the classical equation
\be
\square \ue + m^2 \ue - e^2  A^\mu A_\mu \ue = 0 \,.
\en
The Schr\"odinger equation corresponds to a spin-1 field with mass squared $e^2 \ue^2$ \cite{struyve10}.

\subsection{Temporal gauge}\label{temporal}
In this section, we consider a Bohmian theory for scalar electrodynamics where not all gauge freedom is eliminated. It will appear that we need to deal with the remaining gauge freedom in order to get an adequate Bohmian semi-classical approximation. 

\subsubsection{Classical theory}
The temporal gauge is given by $A_0=0$. It does not completely fix the gauge. There is still a residual gauge symmetry given by the time-independent transformations :
\be
\phi \to \ee^{\ii e \theta} \phi \,, \qquad {\bf A} \to {\bf A} + \bn \theta \,,
\label{tg.00.1}
\en
with ${\dot \theta} = 0$.

In this gauge, the classical equations of motion read
\be
{\ddot \phi} - {\bf D}^2 \phi + m^2 \phi = 0\,, \qquad \square {\bf A} + \bn \bn \cdot {\bf A} = {\bf j} \,, \qquad - \bn \cdot {\dot {\bf A}} = j_0 \,,
\label{tg.00.2}
\en
where ${\bf D} = {\boldsymbol \nabla} - \ii e {\bf A}$, ${\bf j} = \ii e (\phi {\bf D}^*\phi^* - \phi^* {\bf D} \phi)$ and $j_0 = \ii e(\phi^* {\dot \phi} - \phi{\dot \phi}^*)$.

\subsubsection{Bohmian theory}\label{temporalbohmian}
The Schr\"o\-din\-ger equation for $\Psi(\vp,\ca,t)$ is{\footnote{In \cite{kiefer92} this equation is considered in order to study the semi-classical approximation in the context of standard quantum theory.}}
\be
\ii \pa_t \Psi =   \int d^3 x \bigg(-\frac{ \delta^2  }{ \delta \vp^* \delta \vp} + |({\boldsymbol \nabla} - \ii e \ca) \vp|^2  + m^2 |\vp|^2 - \frac{1}{2} \frac{\delta^2}{ \delta \ca^2} + \frac{1}{2} (\bn \times \ca)^2  \bigg)\Psi \,,   
\label{tg.01}
\en
together with the constraint
\be
\bn \cdot \frac{\delta \Psi} {  \delta \ca } + \ii e \left(\vp^* \frac{ \delta \Psi }{\delta \vp^*} -  \vp \frac{ \delta \Psi }{ \delta \vp} \right) = 0 \,,
\label{tg.02}
\en
for the wave functional $\Psi(\vp,\ca,t)$. The constraint expresses the fact that the wave functional is invariant under time-independent gauge transformations, i.e., $\Psi(\vp,\ca)=\Psi(\ee^{\ii e \theta}\vp,\ca +  \bn \theta)$, with $\theta$ time-independent. The constraint is compatible with the Schr\"odinger equation: if it is satisfied at one time, it is satisfied at all times. 

The actual configurations $\phi$ and ${\bf A}$  satisfy \cite{valentini92}:
\be
{\dot \phi} =  \frac{\delta S}{\delta \vp^* }\Bigg|_{\phi,{\bf A}} \,, \qquad {\dot {\bf A}} =  \frac{\delta S}{\delta \ca  }\Bigg|_{\phi,{\bf A}} \,.
\label{tg.03}
\en
These equations are invariant under the time-independent gauge transformations \eqref{tg.00.1} because of the constraint \eqref{tg.02}.

The corresponding second-order equations are
\be
{\ddot \phi} - D^2 \phi + m^2 \phi = - \frac{\delta Q^\Psi}{\delta \vp^*}\Bigg|_{\phi,{\bf A}} \,, \qquad \square {\bf A} + {\boldsymbol \nabla}   {\boldsymbol \nabla} \cdot {\bf A} = {\bf j} + {\bf j}_Q \,,
\label{tg.04}
\en
where
\be
Q^\Psi = - \frac{1}{|\Psi|} \int d^3 x \left( \frac{\delta^2 }{\delta \vp^* \delta \vp} + \frac{1}{2} \frac{\delta^2 }{ \delta \ca^2} \right)|\Psi|
\label{tg.05}
\en
and
\be
{\bf j}_Q  = - \frac{\delta Q^\Psi}{\delta \ca}\Bigg|_{\phi,{\bf A}} \,.
\label{tg.06}
\en
The constraint \eqref{tg.02} further implies that
\be
- {\boldsymbol \nabla} \cdot {\dot {\bf A}} = j_0 \,,
\label{tg.07}
\en
so that, assuming the gauge $A_0 = 0$, \eqref{tg.04} and \eqref{tg.07} can be written as 
\be
D_\mu D^\mu \phi - m^2 \phi = - \frac{\delta Q^\Psi}{\delta \vp^*}\Bigg|_{\phi,{\bf A}} \,, \qquad \pa_\mu F^{\mu \nu} = j^\nu + j^\nu_Q \,,
\label{tg.08}
\en
where $j^\nu_Q = (0,{\bf j}_Q)$.

This Bohmian theory is equivalent to the one formulated using the Coulomb gauge (and hence also to the one formulated using the unitary gauge). To see this, consider the field transformation $\phi,{\bf A} \to \phi',{\bf A}^T,{\bf A}^L$ defined by $ \phi' = \phi \exp(-\ii e \frac{1}{\nabla^2} {\boldsymbol \nabla} \cdot {\bf A})$ and the usual decomposition of ${\bf A}$ into transverse and longitudinal part. In terms of the new variables the wave function is $\Psi'(\vp',\ca^T,\ca^L)=\Psi(\vp,\ca)$. Because of the constraint \eqref{tg.02} we have $\delta \Psi' / \delta \ca^L = 0$ and hence the wave functional does not depend on $\ca^L$. For a solution $\Psi$ to the Schr\"odinger equation \eqref{tg.01}, $\Psi'$ will be a solution to the Schr\"odinger equation \eqref{sq.8} in the Coulomb gauge. (The latter equivalence is discussed in detail in \cite{kim90}.) The guidance equations \eqref{tg.03} also reduce to the ones in the Coulomb gauge. They yield the extra equation $-\bn \cdot \dot {\bf A}^L= j^0$. This means that $j^0$ determines ${\bf A}^L$ up to a time-independent gauge transformation. As such, ${\bf A}^L$ is not an independent dynamical degree of freedom and can be ignored. In the Coulomb gauge there is no field ${\bf A}^L$ and hence the Bohmian formulations are equivalent. (See section \ref{abtemporalbohmian} and \cite{struyve10} for a similar comparison respectively in the case of the Abraham model and the free electromagnetic field.)

\subsubsection{Usual semi-classical approximation}
In the framework of standard quantum theory, there is a natural semi-classical approximation that treats the vector potential classically and the scalar field quantum mechanically. The scalar field is described by a wave functional $\chi(\vp,t)$ which satisfies
\be
\ii \pa_t \chi =   \int d^3 x \bigg(-\frac{ \delta^2  }{ \delta \vp^* \delta \vp} + |{\bf D} \vp|^2  + m^2 |\vp|^2 \bigg)\chi    
\label{tg.0201}
\en
and the electromagnetic field $A^\mu = (0,{\bf A})$ in the temporal gauge satisfies the classical Maxwell equations 
\be
\pa_\mu F^{\mu \nu} = \langle \chi | {\widehat j}^\nu | \chi \rangle \,,
\label{tg.0202}
\en
where
\be
\langle \chi | {\widehat j}^0 | \chi \rangle = \int {\mathcal D} \vp \chi^* {\mathcal C} \chi =  e \int {\mathcal D} \vp \chi^* \left(\vp^* \frac{ \delta \chi }{\delta \vp^*} -  \vp \frac{ \delta \chi }{ \delta \vp} \right) \,,
\nonumber
\en
\be
\langle \chi |{\widehat {\bf j}} | \chi \rangle = \ii  e \int {\mathcal D} \vp |\chi|^2  \left(\vp {\bf D}^*\vp^* - \vp^* {\bf D} \vp\right)\,,
\label{tg.0203}
\en
where the operator ${\mathcal C}$ was defined in \eqref{sq.8.001}.

This theory is consistent since $\pa_\mu \langle \chi |{\widehat j}^\mu| \chi  \rangle = 0$, as a consequence of the Schr\"odinger equation \eqref{tg.0201}. (It is also invariant under the time-independent gauge transformations ${\bf A} \to {\bf A}'={\bf A} + \bn \theta$, $\chi (\vp) \to \chi' (\vp) = \chi (\ee^{-\ii e \theta} \vp)$.)

\subsubsection{Bohmian semi-classical approximation}
A natural guess for a Bohmian semi-classical approximation similar to the usual one is the following. An actual field $\phi$ is introduced that satisfies ${\dot \phi} = ( \delta S/\delta \vp^*)|_{\phi}$, where the wave functional satisfies \eqref{tg.0201}, and the Maxwell equations are $\pa_\mu F^{\mu \nu} = j^\nu$, where $j^\mu$ is the classical expression for the charge current. However, the second-order equation for the Bohmian field is
\be
{\ddot \phi} - D^2 \phi + m^2 \phi = - \frac{\delta Q^\chi}{\delta \vp^*}\Bigg|_{\phi} \,, 
\label{tg.09}
\en
where $Q^\chi = - \frac{1}{|\chi|} \int d^3 x \left( \frac{\delta^2 |\chi|}{\delta \vp^* \delta \vp}\right)$. As a consequence, we have that $\pa_\mu j^\mu = - \ii {\mathcal C} Q^\chi|_{\phi} $ and hence Maxwell's equations imply that ${\mathcal C} Q^\chi =0$ or $Q^\chi=Q^\chi(|\vp|^2)$. This is a constraint on the wave functional that was absent in the usual semi-classical theory. It also seems to be a rather strong condition. It will for example be satisfied if the scalar field evolves classically (i.e., when the right-hand side of \eqref{tg.09} is zero) but it is unclear whether there are other solutions.

We arise to a similar conclusion from a more careful approach trying to derive a semi-classical approximation from the full Bohmian theory. If we want a semi-classical approximation with ${\bf A}$ classical, then we should require that ${\bf j}_Q = 0$ (since we want the wave functional to depend solely on the scalar field and no longer on the vector potential). However, due to the constraint \eqref{tg.02}, we have that ${\boldsymbol \nabla} \cdot {\bf j}_Q|_{\phi,{\bf A}} = \ii {\mathcal C} Q^\Psi|_{\phi,{\bf A}} $ and hence ${\mathcal C} Q^\Psi|_{\phi,{\bf A}} =0$. 

So the conclusion seems to be that if we assume ${\bf A}$ classical, then $\phi$ should also behave classically. This is not surprising since the gauge symmetry implies that the physical (i.e., gauge-invariant) degrees of freedom are some combination of the fields ${\bf A}$ and $\phi$. So one can not just assume ${\bf A}$ classical and keep $\phi$ fully quantum.

A possible way to develop a semi-classical approximation is thus to separate gauge degrees of freedom from gauge-independent ones and assume only some of the latter to be classical. One way to do this is as follows. Writing ${\bf A} = {\bf A}^T + {\bf A}^L$, we can assume ${\bf A}^T$ to behave classically and not ${\bf A}^L$, since ${\bf A}^T$ does not change under a gauge transformation, whereas ${\bf A}^L$ does. We could fully eliminate ${\bf A}^L$ by introducing the field variable $\phi' = \phi \exp(- \ii e \frac{1}{\nabla^2} {\boldsymbol \nabla} \cdot {\bf A})$, which would lead to the full Bohmian theory and the semi-classical approximation of section \ref{cgscacef}. However, we can also formulate a semi-classical approximation by keeping the variable ${\bf A}^L$, but which is still equivalent to the one of section \ref{cgscacef}. The functional Schr\"odinger equation for $\Psi(\vp,\ca^L)$ is
\be
\ii \pa_t \Psi =   \int d^3 x \bigg(-\frac{ \delta^2  }{ \delta \vp^* \delta \vp} +  | [{\boldsymbol \nabla} - \ii e ({\bf A}^T + \ca^L)] \vp|^2  + m^2 |\vp|^2  - \frac{1}{2} \frac{\delta^2}{ \delta \ca^{L2} }  \bigg)\Psi    
\en
and the constraint 
\be
\bn \cdot \frac{\delta \Psi} {  \delta \ca^L} + \ii e \left(\vp^* \frac{ \delta \Psi }{\delta \vp^*} -  \vp \frac{ \delta \Psi }{ \delta \vp} \right) = 0 \,.
\en
The guidance equations are
\be
{\dot \phi} =  \frac{\delta S}{\delta \vp^* }\Bigg|_{\phi,{\bf A}^L} \,, \qquad {\dot {\bf A}^L } =  \frac{\delta S}{\delta \ca^L }\Bigg|_{\phi,{\bf A}^L} \,.
\en
The equation of motion for ${\bf A}^T$ is
\begin{equation}
\square {\bf A}^T  = {\bf j}^T \,.
\end{equation}
Together these equations imply
\be
D_\mu D^\mu \phi - m^2 \phi = - \frac{\delta Q}{\delta \vp^*}\Bigg|_{\phi,{\bf A}^L} \,, \qquad \pa_\mu F^{\mu \nu} = j^\nu + j^\nu_Q
\en
in the temporal gauge, where $j^\nu_Q = (0,{\bf j}_Q)$, with
\be
{\bf j}_Q=  - \frac{\delta Q^\Psi}{\delta\ca^L }\Bigg|_{\phi,{\bf A}^L} \,,
\en
and quantum potential
\be
Q^\Psi = - \frac{1}{|\Psi|} \int d^3 x \left( \frac{\delta^2}{\delta \vp^* \delta \vp} + \frac{1}{2} \frac{\delta^2 }{ \delta \ca^{L2}} \right)|\Psi| \,.
\en
We still have gauge invariance under time independent gauge transformations. This semi-classical approximation is equivalent to the one of section \ref{cgscacef} with the Coulomb gauge. Similarly as in the case of the full Bohmian theories (cf.\ section \ref{temporalbohmian}), this can be shown by applying the field transformation $(\phi,{\bf A}^L) \to (\phi',{\bf A}^L)$ with $\phi' = \phi \exp(-\ii e \frac{1}{\nabla^2} {\boldsymbol \nabla} \cdot {\bf A}^L)$.

\section{Gravity}\label{qg}
\subsection{Canonical quantum gravity}
In canonical quantum gravity, the state vector is a functional of a spatial metric $h_{ij}({\bf x})$ on a 3-dimensional manifold and the matter degrees of freedom, say a scalar field $\varphi({\bf x})$. The wave functional is static and merely satisfies the constraints \cite{kiefer04}:
\be
{\mathcal H} \Psi(h,\vp) = 0 \,,
\en
\be
{\mathcal H}_i \Psi(h,\vp) = 0\,, \quad i=1,2,3 \,.
\en
Their explicit forms are not important here. The latter constraints express the fact that the wave functional is invariant under infinitesimal diffeomorphisms of 3-space (i.e., spatial coordinate transformations). The former equation is the Wheeler-DeWitt equation. It is believed that this equation contains the dynamical content of the theory. However, it is as yet not clear how this dynamical content should be extracted. This is the problem of time \cite{kuchar92,kiefer04}.

In the Bohmian theory, there is an actual 3-metric and a scalar field, whose dynamics depends on the wave function \cite{shtanov96,goldstein04,pinto-neto05a,pinto-neto19}. The dynamics expresses how the Bohmian configuration changes along a succession of 3-dimensional space-like surfaces, which forms a foliation of space-time.{\footnote{The succession of the surfaces is determined by the lapse function and different choices of lapse function lead to a different Bohmian dynamics. This is analogous to the fact that in Minkowski space-time, the Bohmian dynamics (at least in the usual formulation) depends on the choice of a preferred reference frame or foliation.}} Although the wave function is stationary, the Bohmian configuration will change along these surfaces for generic wave functions. This is how the Bohmian theory solves the problem of time. 

The structure of the theory is similar to that of scalar electrodynamics in the temporal gauge, which was discussed in section \ref{temporal}. Namely, in both cases there is a constraint on the wave functional which expresses invariance under infinitesimal gauge transformations: spatial diffeomorphisms in the case of gravity and U(1) transformations in the case of scalar electrodynamics. Therefore we may encounter similar complications in developing a consistent Bohmian semi-classical approximation. Indeed, as we will see below, the Bohmian energy-momentum tensor (for the matter) will not be covariantly conserved and hence can not be used as the source term in the classical Einstein equations. This feature is analogous to what we saw in the case of scalar electrodynamics in the temporal gauge. In that case, the Bohmian charge current was not conserved so that it could not enter as the source in Maxwell's equations. The possible solution to the problem is presumably similar to that in the case of electrodynamics, namely the gauge invariance should be eliminated by working with gauge-invariant degrees of freedom or by choosing a gauge. However, this is a notoriously hard problem in the case of gravity. This problem does not appear in simplified symmetry-reduced models of quantum gravity called mini-superspace models since the spatial diffeomorphism invariance is eliminated. We will discuss such a model in section \ref{mini}.

First, we will consider the problem in formulating a Bohmian semi-classical approximation in more detail. In sections \ref{scgmf} and \ref{scgmp} we will respectively consider a matter field and matter particles. Instead of starting from the Bohmian formulation for canonical quantum gravity, it will be simpler to start from quantized matter on an classical curved space-time. This already allows us to find a natural candidate for the energy-momentum tensor that could be used in the classical Einstein field equations. However, in both cases the natural energy-momentum tensor is not conserved. This suggest that we should start from the full quantum theory with some appropriate elimination of the gauge freedom. In section \ref{bsne} we will consider a Bohmian semi-classical theory in terms of non-relativistic point-particles coupled to Newtonian gravity, which is consistent. For the mini-superspace model, we will derive the semi-classical approximation from the full quantum theory.

\subsection{Semi-classical gravity: matter field}\label{scgmf}
The formulation of quantum field theory on a classical curved space-time in the functional Schr\"odinger picture was detailed in \cite{halliwell91a,long98}. We assume that the space-time manifold ${\mathcal M}$ is globally hyperbolic so that it can be foliated into space-like hypersurfaces. ${\mathcal M}$ is then diffeomorphic to ${\mathbb R} \times \Sigma$, with $\Sigma$ a 3-surface. We choose coordinates $x^\mu=(t,{\bf x})$ such that the time coordinate $t$ labels the leaves of the foliation and ${\bf x}$ are coordinates on $\Sigma$. In terms of these coordinates the space-time metric and its inverse can be written as 
\be
g_{\mu \nu}= 
\begin{pmatrix}
N^2 - N_i N^i & -N_i \\
-N_i & - h_{ij}
\end{pmatrix} \,,
\qquad
g^{\mu \nu}= 
\begin{pmatrix}
\frac{1}{N^2} & \frac{-N^i}{N^2} \\
\frac{-N^i}{N^2} &   \frac{N^iN^j}{N^2}- h^{ij}
\end{pmatrix} \,,
\en
where $N$ is the lapse function and $N_i = h_{ij}N^j$ are the shift functions. $h_{ij}$ is the induced Riemannian metric on the leaves of the foliation. The unit vector field normal to the leaves is $n^\mu=(1/N, -N^i/N)^T$. 

For a real mass-less scalar field, the Schr\"odinger equation for this space-time background reads
\be
\ii \frac{\pa \Psi}{\pa t} = \int_{\Sigma} d^3 x \left(N {\widehat {\mathcal H}} + N^i {\widehat {\mathcal H}}_i \right) \Psi \,, 
\label{scg.05}
\en
where $\Psi$ is a functional on the space of fields $\vp$ on $\Sigma$ and
\be
{\widehat {\mathcal H}} = \frac{1}{2} {\sqrt h} \left(-\frac{1}{h} \frac{\de^2}{\de \vp^2}  + h^{ij} \pa_i \vp \pa_j \vp\right)\,,
\label{scg.06}
\en
\be
{\widehat {\mathcal H}}_i =- \frac{\ii}{2} \left(\pa_i \vp \frac{\de}{\de \vp}  + \frac{\de}{\de \vp} \pa_i \vp \right)\,.
\label{scg.07}
\en

This equation describes the action of the classical metric onto the quantum field. The usual way to introduce a back-reaction is by using the expectation value of the energy-momentum tensor operator as the source term in Einstein's field equations, i.e.,
\be
G_{\mu \nu} (g) = \langle \Psi | {\widehat T}_{\mu \nu} (\vp,g) |\Psi\rangle \,.
\label{scg.08}
\en
This equation is consistent since $\nabla^\mu \langle \Psi | {\widehat T}_{\mu \nu} (\vp,g) |\Psi\rangle = 0$ because of \eqref{scg.05}.

In the Bohmian theory, the actual scalar field $\phi$ satisfies the guidance equation
\be
{\dot \phi} = \frac{N}{\sqrt{h}} \frac{\de S}{\de \vp}\Bigg|_{\phi} + N^i \pa_i \phi 
\en
or, equivalently,
\be
n^\mu \pa_\mu \phi = \frac{1}{\sqrt{h}} \frac{\de S}{\de \vp}\Bigg|_{\phi} \,.
\en
This dynamics depends on the foliation (unlike the wave function dynamics). That is, different foliations will lead to different evolutions of the scalar field.

Taking the time derivative of this equation, we obtain 
\be
\nabla_\mu \nabla^\mu \phi = - \frac{1}{\sqrt{ -g}} \frac{\de Q^\Psi}{\de \vp}\Bigg|_{\phi} \,,
\label{scg}
\en
with
\be
Q^\Psi (\vp,t) =   -\frac{1}{2 |\Psi(\vp,t)|}  \int d^3x \frac{ N({\bf x},t)}{\sqrt{ h({\bf x},t)}}   \frac{\de^2 |\Psi(\vp,t)|}{\de \vp^2({\bf x})}  
\en
the quantum potential. 

To find the energy-momentum tensor for the Bohmian field, we consider the action 
\be
S_B = \frac{1}{2} \int d^4 x {\sqrt{ -g}} \left( g^{\mu \nu} \pa_\mu \phi  \pa_\nu \phi\right) - \int dt Q^\Psi(\phi) 
\en
from which the equation of motion \eqref{scg} can be derived. We can then use the usual definition of the energy-momentum tensor 
\be
T^{\mu \nu} = - \frac{2}{\sqrt{ -g}} \frac{\de S_B}{\de g_{\mu \nu}}
\label{scg.5}
\en
to obtain
\be
T^{\mu \nu} =  \pa^\mu \phi \pa^\nu  \phi  - \frac{1}{2} g^{\mu \nu} g^{\al \beta} \pa_\al \phi \pa_\beta \phi + T_Q^{\mu \nu} \,,
\label{scg.5.1}
\en
which has the form of the classical expression with a quantum contribution
\be
T_Q^{\mu \nu} =  \frac{2}{\sqrt{ -g}} \frac{\de \int dt Q^\Psi(\phi)}{\de g_{\mu \nu}} = - \frac{1}{2h|\Psi|} \frac{\de^2 |\Psi|}{\de \vp^2}\Bigg|_{\phi} \left( n^\mu n^\nu + \de^\mu_i\de^\nu_j h^{ij}  \right)\,.
\en
The expression \eqref{scg.5.1} seems to be the natural one for the Bohmian energy-momentum tensor, which could be used in the classical Einstein field equations. However, by taking the covariant derivative, we find
\be
\nabla_\mu T^{\mu \nu} = -  \frac{1}{\sqrt{ -g}}   \frac{\de Q^\Psi}{\de \vp} \pa^\nu \vp   + \nabla_\mu T_Q^{\mu \nu}\,,
\en
which is generically different from zero, so that Einstein's equations would not be consistent with this energy-momentum tensor as matter source. Note that even without the contribution $T_Q^{\mu \nu}$, the energy-momentum tensor is not conserved. 

The energy-momentum tensor was derived here in the semi-classical context. In a more precise derivation, we should start from the Bohmian theory of quantum gravity and consider the modified Einstein field equations \cite{duerr20b} to find the energy-momentum tensor for the matter field. However, unless the spatial diffeomorphism invariance is eliminated, this would not help in finding a consistent semi-classical approximation. Once this invariance is eliminated one can expect a further contribution to the energy-momentum tensor that makes it conserved.

\subsection{Semi-classical gravity: matter particles}\label{scgmp}
A similar consistency problem arises when point-particles are considered. To illustrate this, we will consider a single Klein-Gordon particle. (The extension to many particles is straightforward.) The wave equation in an external gravitational field is
\be
g^{\mu \nu} \nabla_\mu \nabla_\nu \psi(x) + m^2 \psi(x) = 0\,.
\en
The guidance equation for a point-particle with space-time coordinates $X^\mu$ is
\be
\frac{d X^\mu}{ds} = - \frac{1}{m} \nabla^\mu S|_{X} \,.
\en
Note that this guidance equation determines a particular parameterization of the world-line. The proper-time $\tau$ is given by $d\tau = \sqrt{1+2Q^\psi(X(s))/m} ds$, where
\be
Q^\psi = \frac{1}{2mR} \nabla^\mu \nabla_\mu R \,, \qquad R=|\psi| \,,
\en
is the quantum potential. The parameterization by $s$ will be most convenient to express the energy-momentum tensor. (As is well-known, even in the case of flat space-time the Bohmian trajectories are not always time-like, not-even when restricting to positive energies \cite{kyprianidis85,holland93b}. However, this problem is unrelated to finding a consistent semi-classical theory. Instead of the Klein-Gordon theory one could also consider the Dirac theory, where the trajectories are time-like. However, this would entail unnecessary complications.)

The guidance equation implies the following second-order equation:
\be
\frac{d^2 X^\mu}{ds^2} + \Gamma^\mu_{\al \beta} \frac{d X^\al}{ds}\frac{d X^\beta}{ds} = \frac{1}{m} \pa^\mu Q^\psi(X) \,.
\en
This equation can be derived from the following action
\be
S_B=-\int ds \left(\frac{m}{2} g_{\mu \nu} (X) \frac{d X^\mu}{ds}\frac{d X^\nu}{ds} + Q^\psi(X)\right)\,.
\en
Using again the definition \eqref{scg.5} for the energy-momentum tensor, we obtain
\begin{align}
  T^{\mu \nu}(x)  &= \frac{1}{\sqrt{-g(x)}}  \int ds \delta(x - X) m \frac{d X^\mu}{ds}\frac{d X^\nu}{ds} + T^{\mu \nu}_Q(x)   \nonumber\\
  &= \frac{1}{\sqrt{-g(x)}}\frac{1}{m} \nabla^\mu S(x) \nabla^\nu S(x) \int ds \delta(x - X) + T^{\mu \nu}_Q(x)  \,, 
\end{align}
which is the classical expression with a quantum contribution
\be
T^{\mu \nu}_Q(x) = \frac{2}{\sqrt{-g(x)}} \frac{\de \int ds Q^\psi }{\de g_{\mu \nu}(x)} .
\en
This expression could be further worked out. However, just as in the case of a matter field, the energy-momentum tensor is not conserved but gives
\be
  \nabla_\mu T^{\mu \nu}(x) =    \frac{1}{\sqrt{-g(x)}} \partial^\nu Q^\psi(x)\int ds \delta(x - X) + \nabla_\mu T^{\mu \nu}_Q(x) \,.
\en

In the case of the Abraham model, which also conserns point-particles, no inconsistencies were encountered. But the important different is that in the case of the Abraham model, as noted at the end of section \ref{abrahamtempsca}, a gauge transformation does not act on the particles, but just on the electromagnetic field, while in the case of gravity a gauge transformation (i.e.\ diffeomorphism transformation) acts on both the particle and the metric field. Presumably, that is the reason why a consistent semi-classical approximation does not follow directly in the latter case.

\subsection{Bohmian analogue of the Schr\"odinger-Newton equation}\label{bsne}
In the previous section, we explained the problem in finding a consistent Bohmian semi-classical approximation by coupling classical gravity to relativistic point-particles. If we consider the Newtonian limit of gravity and the non-relativistic limit for the quantum matter, we obtain a consistent Bohmian semi-classical theory.  

Without explicitly performing these limits to the semi-classical theory outlined in the previous section (one could follow the analysis of \cite{suarez15a}), the resulting theory is given by the Schr\"odinger equation
\be
\ii \pa_t \psi({\bf x},t) = \left[ - \frac{1}{2m} \nabla^2  +m \Phi({\bf x},t) \right] \psi({\bf x},t)\,,
\label{sn.1}
\en
where $\Phi$ is the gravitational potential, which satisfies the Poisson equation
\be
\nabla^2 \Phi({\bf x},t) = 4 \pi Gm \delta({\bf x} - {\bf X}(t)) \,,
\label{sn.5}
\en
and where ${\bf X}$ is the position of the Bohmian particle, which satisfies the guidance 
equation
\be
{\dot {\bf X}}(t) = {\bf v}^\psi({\bf X}(t),t) \,.
\label{sn.4}
\en
So the Bohmian particle acts as the gravitational source. Solving for the potential, one can write \eqref{sn.1} as
\be
\ii \pa_t \psi({\bf x},t) =  \left[  - \frac{1}{2m} \nabla^2  -  \frac{G m^2}{|{\bf x}- {\bf X}(t)|}  \right]\psi({\bf x},t)\,.
\label{sn.6}
\en

The generalization to many particles is straightforward. In this case, the gravitational potential generated by the Bohmian particles is 
\be
\Phi({\bf x},t) = - G\sum_k m_k \frac{1}{|{\bf x}- {\bf X}_k(t)|} \,,
\en
so that the potential in the many-particle Schr\"odinger equation is given by 
\be
V(x) = \sum_k m_k \Phi({\bf x}_k,t) \,.
\en

This semi-classical approximation is similar to the Schr\"odinger-Newton equation \cite{diosi84}, which can be derived from \eqref{scg.05} and \eqref{scg.08} \cite{bahrami14}, and which is given by \eqref{sn.1} together with
\be
\nabla^2 \Phi = 4\pi Gm |\psi|^2 \,.
\label{sn.2}
\en
So in this case the mass density $m|\psi|^2$ acts as the gravitational source. Solving for the potential, the following non-linear Schr\"odinger equation
\be
\ii \pa_t \psi({\bf x},t) =  \left[- \frac{1}{2m} \nabla^2  - G m^2 \int d^3y \frac{|\psi({\bf y},t)|^2}{|{\bf x}- {\bf y}|}\right] \psi({\bf x},t)
\label{sn.3}
\en
is obtained.

While the Bohmian theory is now consistent, it should be noted that it might not directly follow from a Bohmian quantum gravity with point-particles, because of the lessons that were drawn in the case of the Abraham model in section \ref{scaqam}. For example, in the case of the Coulomb gauge, we found (slightly) different theories by starting from a natural guess for a semi-classical theory (cf.\ section \ref{ascaam}) or by deriving it from the full Bohmian theory for the Abraham model (cf.\ section \ref{asca}). In the former case, the Coulomb potential that appears in the Schr\"odinger equation was generated by the Bohmian charges, while in the latter case it is not. In the latter case the Coulomb potential just takes the usual form as a function on configuration space. In the case of gravity, we have similarly merely guessed the semi-classical approximation rather than having derived it from a Bohmian quantum gravity. This being said, the difference between the two version in the case of the Abraham model seem minor. Similarly, the Bohmian analogue of the Schr\"odinger-Newton equation might form a good semi-classical approximation. (It could perhaps also be seen as an alternative to collapse models. A model similar in spirit was studied in \cite{laloe15}.)

\subsection{Mini-superspace model}\label{mini}
In this section, we consider a symmetry-reduced model of quantum gravity where homogeneity and isotropy are assumed. In this model, the spatial diffeomorphism invariance is eliminated and we can straightforwardly develop a Bohmian semi-classical approximation. 

In the classical mini-superspace model, the universe is described by the Friedmann-Lema\^itre-Robertson-Walker (FLRW) metric 
\be
\ud s^2 =  N(t)^2 \ud t^2 - a(t)^2 \ud \Omega^2_3 \,,
\label{ms.01}
\en
where $N$ is the lapse function, $a=\ee^{\al}$ is the scale factor{\footnote{The reason for introducing the variable $\al$ is that it is unbounded, unlike the scale factor, which satisfies $a>0$.}}  and $\ud \Omega^2_3$ is the metric on 3-space with constant curvature $k$. Assuming matter that is described by a homogeneous scalar field $\phi$, the Lagrangian is \cite{halliwell91b,vink92}
\be
L = N \ee^{3\al} \left[ - \ka \left( \frac{{\dot \al}^2 }{2N^2} +  V_G(\al)\right)  + \frac{{\dot \phi}^2 }{2N^2}  - V_M(\phi)\right]\,,
\label{ms.02}
\en
where $\ka = 3/4\pi G$.  
\be
V_G(\al)= - \frac{1}{2} k \ee^{-2\al} + \frac{1}{6} \Lambda 
\label{ms.03}
\en
is the gravitational potential, with $\Lambda$ the cosmological constant, and $V_M$ is the potential for the matter field. The corresponding equations of motion are, after imposing the gauge $N=1$:{\footnote{The theory is time-reparamaterization invariant. Solutions that differ only by a time-reparameterization are considered physically equivalent. Choosing the gauge $N=1$ corresponds to a particular time-parameterization.}}
\begin{align}
& \frac{1}{2} {\dot \al}^2 = \frac{1}{\ka}\left(\frac{1}{2} {\dot \phi}^2 +V_M(\phi)\right) + V_G(\al) \,,\label{ms.031}\\
& \ddot \phi + 3 \dot \al \dot \phi + \pa_\phi V_M(\phi) = 0 \,.
\label{ms.04}
\end{align}
(The second-order equation for $\al$ which arises from variation with respect to $\al$ is redundant since it can be derived from the other two equations.)

Using the canonical momenta
\be
\pi_N =0  \,, \qquad \pi_\al = - \ka \ee^{3\al} \frac{\dot \al}{N} \,, \qquad \pi_\phi = \ee^{3\al} \frac{\dot \phi}{N} \,,
\label{ms.05}
\en
we can pass to the Hamiltonian formulation. This leads to the Hamiltonian constraint (which is just eq.\ \eqref{ms.031})
\be
- \frac{1}{2 \ka \ee^{3\al}}  \pi^2_\al + \frac{1}{2\ee^{3\al} } \pi^2_\phi  +  \ee^{3\al} \left[ \ka V_G(\al) + V_M(\phi) \right] = 0 \,.
\label{ms.06}
\en
Quantization yields the Wheeler-DeWitt equation:
\be
({\widehat H}_G + {\widehat H}_M) \psi(\vp,\ua) = 0 \,,
\label{ms.07}
\en
where
\be
{\widehat H}_G = \frac{1}{2\ka \ee^{3\ua}} \pa^2_\ua + \ka \ee^{3\ua}V_G(\ua) \,, \qquad {\widehat H}_M = - \frac{1}{2\ee^{3\ua}} \pa^2_\vp + \ee^{3\ua}V_M(\vp) \,.
\label{ms.08}
\en

In the corresponding Bohmian theory \cite{vink92}, the FLRW metric (of the form \eqref{ms.01}) and the scalar field satisfy the guidance equations
\be
\dot \al = -  \frac{N}{\ka\ee^{3\al}} \pa_\ua S\big|_{\vp = \phi, \ua =\al} \,, \qquad \dot \phi= \frac{N}{\ee^{3\al}} \pa_\vp S\big|_{\vp = \phi, \ua =\al} \,, 
\label{ms.09}
\en 
where $N(t)$ is an arbitrary lapse function.{\footnote{Just as the classical theory, the Bohmian theory is time-reparameterization invariant. This is a special feature of mini-superspace models \cite{acacio98,falciano01}. As mentioned before, for the usual formulation of the Bohmian dynamics for the Wheeler-DeWitt theory of quantum gravity, a particular space-like foliation of space-time or, equivalently, a particular choice of ``initial'' space-like hypersurface and lapse function, needs to be introduced. Different foliations (or lapse functions) yield different Bohmian theories.}} In the gauge $N=1$, these equations imply 
\begin{align}
& \frac{1}{2} {\dot \al}^2 = \frac{1}{\ka}\left(\frac{1}{2} {\dot \phi}^2 +V_M(\phi) + Q^\psi_M (\phi,\al) \right) + V_G (\al)  + Q^\psi_G  (\phi,\al) \,,\label{ms.09.1}\\
& \ddot \phi + 3 \dot \al \dot \phi + \pa_\vp (V_M + Q^\psi_M + \ka Q^\psi_G)\big|_{\vp=\phi,\ua = \alpha} = 0 \,,
\label{ms.09.2}
\end{align}
where 
\be
Q^\psi_G = \frac{1}{2\ka^2 \ee^{6\ua} } \frac{\pa^2_\ua |\psi|}{|\psi|} \,, \qquad Q^\psi_M = - \frac{1}{2 \ee^{6\ua} } \frac{\pa^2_\vp |\psi|}{|\psi|}
 \,.
\label{ms.10}
\en

We will now look for a semi-classical approximation where the scale factor behaves approximately classical. In order to do so, we assume again the gauge $N=1$ and we consider the conditional wave function $\chi(\vp,t) = \psi (\vp,\al(t))$, given a set of trajectories $(\al(t),\phi(t))$. Using 
\be
\pa_t \chi (\vp,t) = \pa_\ua \psi(\vp,\ua) \big|_{\ua = \al(t)} \dot \al(t) \,,
\label{ms.11}
\en
we can write
\be
\ii \pa_t \chi =  {\widehat H}_M \chi + I \,,
\label{ms.12}
\en
where{\footnote{To obtain this equation, note that $\pa^2_\ua \psi = [ (\pa_\ua S)^2 + \ii\pa^2_\ua S  + \pa^2_\ua |\psi|/|\psi|]\psi +2\ii \pa_\ua S \pa_\ua \psi$, so that $\pa^2_\ua \psi|_{\ua = \al(t)} = [ (\pa_\ua S)^2 + \ii\pa^2_\ua  + \pa^2_\ua |\psi|/|\psi|]|_{\ua = \al}\chi +2\ii  \pa_\ua S|_{\ua = \al} \pa_t \chi / \dot \al$. Using this equation together with \eqref{ms.07} we obtain \eqref{ms.12}.}}
\be
I = \frac{1}{\dot \al}\ii \pa_t \chi \left( \dot \al + \frac{1}{ \ka\ee^{3\al}} \pa_\ua S \big|_{\ua =\al}\right) + \frac{1}{2\ka\ee^{3\al}}\left[ (\pa_\ua S)^2 + \ii\pa^2_\ua S\right]\Big|_{\ua =\al}\chi  +\ka \ee^{3\al} (V_G + Q^\psi_G)  \Big|_{\ua = \al} \chi \,.
\label{ms.13}
\en
When $I$ is negligible (up to a real time-dependent function times $\chi$), \eqref{ms.12} becomes the Schr\"odinger equation for a homogeneous matter field in an external FLRW metric. We can further assume the quantum potential $Q^\psi_G$ to be negligible compared to other terms in eq.\ \eqref{ms.09.1}. As such, we are led to the semi-classical theory:
\begin{align}
& \ii \pa_t \chi =  {\widehat H}_M \chi \,,\label{ms.14}\\
& \dot \phi=   \frac{1}{\ee^{3\al}} \pa_\vp S\big|_{\vp = \phi} \,,\label{ms.15}\\
& \frac{1}{2} {\dot \al}^2 = \frac{1}{\ka}\left(\frac{1}{2} {\dot \phi}^2 +V_M(\phi) +Q^\chi_M(\phi,\al)\right) + V_G(\al) \equiv  - \frac{1}{\ka\ee^{3\al}} \pa_t S\big|_{\vp = \phi}  + V_G(\al)  \,. \label{ms.16}
\end{align}

Let us now consider when the term $I$ will be negligible. The quantity in brackets in the first term would be zero when evaluated for the actual trajectory $\phi(t)$ (because of the guidance equation for $\al$). As such, the first term will be negligible if the actual scale factor evolves approximately independently of the scalar field. The second term will be negligible if $S$ varies slowly with respect to $\ua$ or if the term in square brackets is approximately independent of $\vp$. In the latter case, the second term becomes a time-dependent function times $\chi$, which can be eliminated by changing the phase of $\chi$. Similarly, if $Q^\psi_G \ll V_G$ then the third term also becomes a time-dependent function times $\chi$. 

In the usual semi-classical approximation, one has \eqref{ms.14} and
\be
\frac{1}{2} {\dot \al}^2 = \frac{1}{\ka \ee^{3\al}}\langle \chi| {\widehat H}_M(\al)| \chi \rangle + V_G(\al) \,,
 \label{ms.17}
\en
with $\chi$ normalized to one (which is the analogue of \eqref{0.001} and \eqref{0.002} for mini-superspace). In the next section, we will compare this approximation with the Bohmian one for a particular example. It will appear that the latter gives better results than the usual approximation. (Note that Vink himself, in his seminal paper \cite{vink92} which introduces the Bohmian theory for quantum gravity, considers a derivation of the usual semi-classical approximation, rather than the Bohmian one. But he hinted on the Bohmian semi-classical approximation in \cite{kowalski-glikman90}.)

\subsection{Example in mini-superspace}
In this section, we will work out a simple example to compare the Bohmian and usual semi-classical approximations to the full Bohmian result. 

We put $\kappa=1$ and assume $V_G=V_M =0$. It will also be useful to introduce the time parameter $\tau$, defined by $d\tau \ee^{3\al} = dt$ (which corresponds to choosing $N=\ee^{3\al}$ instead of $N=1$). Derivatives with respect to $\tau$ will be denoted by primes. In this way, the classical equations \eqref{ms.031} and \eqref{ms.04} reduce to
\be
\al'^2 = \phi'^2 \,, \qquad \phi''=0 \,.
\label{ems.1}
\en
The possible solutions are
\be
\al = c_1 \tau + c_2\,, \qquad \phi = \pm c_1 \tau + c_3 \,,
\label{ems.2}
\en
where $c_i$, $i=1,2,3$, are constants. In the case $c_1=0$, the scale factor is constant and we have Minkowski space-time. If $c_1 > 0$ the universe starts from a big bang and keeps expanding forever. If $c_1 < 0$ the universe contracts until a big crunch. If $c_1 \neq0$, the corresponding paths in $(\phi,\al)$-space are given by
\be
\al = \pm \phi + c \,,
\en
with $c$ constant. 

\subsubsection{Full Bohmian analysis}
In the full quantum case, we have the Wheeler-DeWitt equation
\be
(\pa^2_\ua - \pa^2_\vp) \psi(\vp,\ua) =0
\label{ems.3}
\en 
and guidance equations
\be
\al' = -\pa_\ua S\big|_{\vp=\phi,\ua = \alpha} \,, \qquad \phi' = \pa_\vp S\big|_{\vp=\phi,\ua = \alpha} \,.
\label{ems.4}
\en

For the wave function
\be
\psi_R(\vp,\ua) = \exp \left[\ii u (\vp -\ua) - \frac{( \vp - \ua )^2}{4\si^2}   \right] \,,
\label{ems.5}
\en
the guidance equations read $\al' =  \phi' = u$, so that we have only classical solutions 
\be
\al = u \tau + c_1 \,, \qquad \phi = u \tau + c_2 
\label{ems.6}
\en
or
\be
\al = \phi + c \,.
\label{ems.6.01}
\en
See fig.\ \ref{rightmoving} for some trajectories.

Similarly, for the wave function
\be
\psi_L(\vp,\ua) = \exp \left[ - \ii v (\vp + \ua) - \frac{( \vp + \ua )^2}{4\si^2}  \right]
\label{ems.7}
\en
the solutions are also classical: 
\be
\al = v \tau + c_1 \,, \qquad \phi = -v \tau + c_2 \,, 
\label{ems.7.01}
\en
or 
\be
\al = - \phi + c \,, 
\label{ems.7.02}
\en
see fig.\ \ref{leftmoving}. 

\begin{figure}[t]
\centering
\begin{minipage}[b]{0.45\linewidth}
\includegraphics[width=0.9\linewidth]{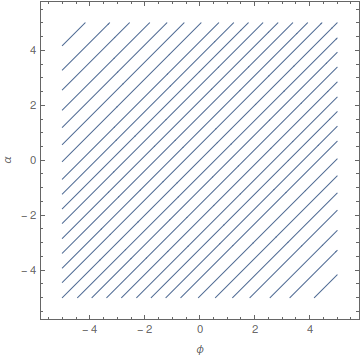}
\caption{Some trajectories for $\Psi_R$, $\si =1$.}
\label{rightmoving}
\end{minipage}
\quad
\begin{minipage}[b]{0.45\linewidth}
\includegraphics[width=0.9\linewidth]{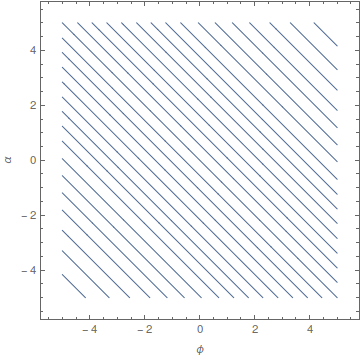}
\caption{Some trajectories for $\Psi_L$, $\si =1$.}
\label{leftmoving}
\end{minipage}
\end{figure}
Consider now the superposition $\psi = \psi_R + \psi_L$ and assume $v > u \gg 0$. For $\ua \to \pm \infty$ the wave functions $\psi_R$ and $\psi_L$ are non-overlapping functions of $\vp$ so that, asymptotically, the Bohmian dynamics is either determined by $\psi_R$ or $\psi_L$. This means that asymptotically, i.e., for $\ua \to \pm \infty$, we have classical motion, given either by \eqref{ems.6.01} or \eqref{ems.7.02}. Some trajectories are plotted in figs.\ \ref{superposition} and \ref{superposition2}. (Trajectories for the case $u=-v$ were plotted in \cite{colistete00}.) We see that trajectories starting from the ``left'', i.e.\ $\phi < 0$, for $\al \to -\infty$ will end up on the left, while trajectories that start from the ``right'', i.e.\ $\phi > 0$, might either end up moving to the left or to the right. (If $u=v$, then trajectories starting on the left stay on the left and trajectories starting on the right stay on the right.) Some trajectories are closed. They correspond to cyclic universes (which oscillate between a minimum and maximum scale factor), see fig.\ \ref{superposition2}.

\begin{figure}[t]
\centering
\begin{minipage}[b]{0.45\linewidth}
\includegraphics[width=0.9\linewidth]{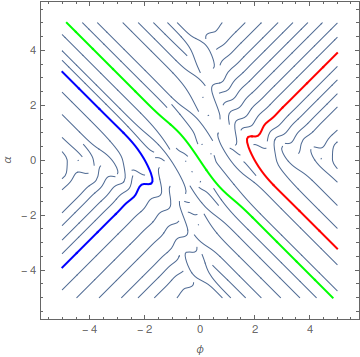}
\caption{Some trajectories (or pieces thereof) for $\Psi_R + \Psi_L$, $u = 1$, $v=5$, $\si =1$. Three trajectories are highlighted which illustrate the possible behaviors for trajectories which asymptotically go like $\al=\pm \phi$.}
\label{superposition}
\end{minipage}
\quad
\begin{minipage}[b]{0.45\linewidth}
\includegraphics[width=0.9\linewidth]{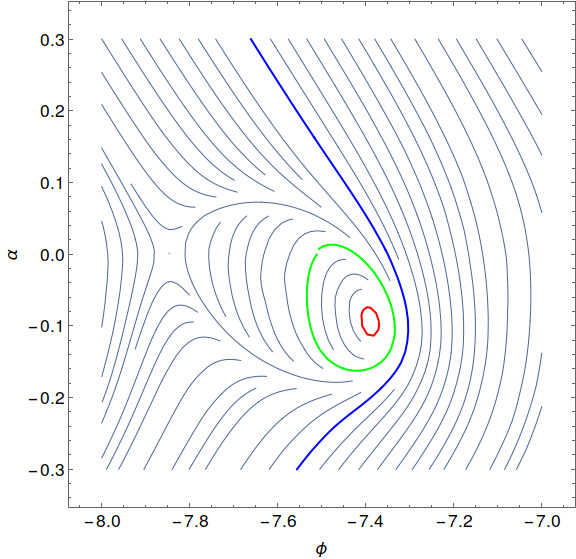}
\caption{Some trajectories (or pieces thereof) for $\Psi_R + \Psi_L$, $u = 1$, $v=5$, $\si =1$. Three trajectories are highlighted. Two of these correspond to closed loops, which correspond to cyclic universes. }
\label{superposition2}
\end{minipage}
\end{figure}

For trajectories with classical asymptotic behavior, there is possible non-classical behavior in the region of overlap, where $\al \approx 0$. For trajectories starting on the left there is a transition from $\al = u \phi$ to $\al = -v \phi$ (which is impossible classically). For some trajectories starting on the right, there is the opposite transition.

Note that there is no natural measure on the set of trajectories \cite{falciano15}, so we can not make any probabilistic statements like about the probability for a trajectory to move from left to right. 

\subsubsection{Usual semi-classical approximation}
Let us first consider the usual semi-classical approximation and compare it to the full Bohmian theory. Eqs.\ \eqref{ms.14} and \eqref{ms.17} reduce to
\be
\ii \pa_\tau \chi = {\widehat h}_M \chi = - \frac{1}{2} \pa^2_\vp \chi \,, 
\label{ems.8}
\en 
\be
 \al'^2 =  2\langle \chi| {\widehat h}_M | \chi\rangle = - \langle  \chi|\pa^2_\vp  |\chi\rangle \,.
\label{ems.9}
\en 
The wave equation corresponds to that of a single particle of unit mass in one dimension. Hence, $\langle  \chi|\pa^2_\vp  |\chi\rangle$ is time-independent and we have that $\al'$ is constant.

The initial wave function which we will use in the semi-classical approximation is given by the conditional wave function $\chi(\vp,\tau_0) = {\mathcal N} \psi(\vp,\al(\tau_0))$ at some time $\tau_0$, with ${\mathcal N}$ a normalization constant. The conditional wave functions corresponding to $\psi_R$ and $\psi_L$ are
\begin{align}
\chi_R(\vp,\tau_0) &= {\mathcal N}_R \psi_R(\vp,\al_0) = (2\pi \sigma^2)^{-1/4} \exp \left[  \ii u (\vp -\al_0)- \frac{( \vp -\al_0 )^2}{4\si^2}  \right] \,,\nonumber\\
\chi_L(\vp,\tau_0) &= {\mathcal N}_L \psi_L(\vp,\al_0) = (2\pi \sigma^2)^{-1/4} \exp \left[ - \ii v (\vp + \al_0) - \frac{( \vp + \al_0 )^2}{4\si^2}  \right]\,,
\label{ems.10}
\end{align}
where $\al_0 = \al(\tau_0)$. $\chi_R$ is a Gaussian packet centered around $\al_0$ and average momentum $u$, while $\chi_L$ is a Gaussian packet centered around $-\al_0$ and average momentum $-v$. The solutions to the Schr\"odinger equation \eqref{ems.8}, with these conditional wave functions as initial conditions, are given by \cite{holland93b}:
\begin{align}
\chi_R(\vp,\tau) &= [2\pi s^2({\bar \tau})]^{-1/4} \exp \left[ \ii u (\vp_R -  u {\bar \tau}/2) - (\vp_R -  u {\bar \tau})^2/4s({\bar \tau})\si\right]\,,\nonumber\\
\chi_L(\vp,\tau) &= [2\pi s^2({\bar \tau})]^{-1/4} \exp \left[- \ii v (\vp_L +  v {\bar \tau}/2) - (\vp_L + v {\bar \tau})^2/4s({\bar \tau})\si\right]\,,
\label{ems.11}
\end{align}
where 
\be
\vp_R = \vp - \al_0 \,, \quad \vp_L = \vp + \al_0    \,, \quad    {\bar \tau} = \tau - \tau_0     \,, \quad   s({\bar \tau}) = \si (1+\ii {\bar \tau}/2 \si^2) \,.
\label{ems.11.1}
\en 
In the following, we assume that
\begin{itemize}
\item
$\tau_0$ is small enough, so that $\al_0 < 0$ and $|\al_0| \gg \si$ (so that $\chi_R$ and $\chi_L$ have little overlap initially),
\item
$\al'(\tau_0)>0$ (\eqref{ems.9} determines $\al'$ only up to a sign),
\item
$v > u \gg 0$,
\item
$1 \ll \si u$.
\end{itemize}

The expectation value $\langle \chi|{\widehat h}_M |\chi\rangle$ is time-independent. So we can calculate it at time $\tau_0$. We have that 
\be
2 \langle\chi_R| {\widehat h}_M |\chi_R\rangle = u^2 + \frac{1}{4\si^2} \,.
\label{ems.12}
\en 
Hence $\al' = \sqrt{u^2 + \frac{1}{4\si^2}}$. Since $1 \ll \si u$, we have that $\al' \approx u$, so that we approximately obtain the classical solution for the scale factor. 

Similarly, for $\chi_L$, we have that  
\be
2 \langle \chi_L|{\widehat h}_M|\chi_L \rangle = v^2 + \frac{1}{4\si^2} 
\label{ems.13}
\en 
and since $1 \ll \si u < \si v$, we have $\al' \approx v$. 

Now consider the superposition $\chi= (\chi_R + \chi_L)/\sqrt{2}$. Since $\chi_R$ and $\chi_L$ have approximately negligible overlap initially, because $|\al_0| \gg \si$, this state is approximately normalized to one and
\be
2 \langle \chi|{\widehat h}_M |\chi\rangle \approx   \langle \chi_R|{\widehat h}_M |\chi_R\rangle + \langle\chi_L| {\widehat h}_M |\chi_L\rangle = \frac{1}{2}(u^2 + v^2) + \frac{1}{4\si^2} \,.
\label{ems.14}
\en
Hence, for $1 \ll \si u$, we have that  $\al' \approx \sqrt{(u^2 + v^2)/2}$. As such, the semi-classical approximation is very close to the exact result, given that $u\approx v$. But if $u$ is very different from $v$, the semi-classical approximation is not so good. In particular, we do not get the asymptotic behavior that $\al' =u$ or  $\al' =v$ for early or late times (i.e., $\tau \to \tau_0$ or $\tau \to \infty$).

\subsubsection{Bohmian semi-classical approximation}
The equations of motion in the Bohmian semi-classical approximation are
\be
\ii \pa_\tau \chi = - \frac{1}{2} \pa^2_\vp \chi \,, 
\label{ems.15}
\en 
\be
\phi' = \pa_\vp S\big|_{\vp=\phi} \,,
\label{ems.16}
\en 
\be
\al'^2 =  \phi'^2 - \frac {\pa^2_\vp |\chi|}{|\chi|}\Bigg|_{\vp=\phi} \equiv -2\pa_\tau S\big|_{\vp=\phi} \,.
\label{ems.17}
\en 
For the state $\chi_R$, the solutions of the guidance equation are \cite{holland93b}:
\be
\phi - \al_0 = u(\tau - \tau_0) + (\phi_0 - \al_0) \frac{|s(\tau - \tau_0))|}{\si} 
\label{ems.18}
\en 
or
\be
\phi_R = u {\bar \tau} + \phi_{R,0} \frac{|s({\bar \tau})|}{\si} \,,
\label{ems.18.1}
\en
where $\phi_0 = \phi(\tau_0)$ and $\phi_{R,0} = \phi_R(\tau_0) = \phi_0 - \al_0$ and the notation \eqref{ems.11.1} is used. With the same assumptions about the constants as in the previous section and taking $|\phi_{R,0}|  \lesssim \si$, i.e., that the initial value $\phi_{R,0}$ does not lie too far outside the bulk of the Gaussian packet, we have that 
\be
\phi_R \approx u {\bar \tau} + \phi_{R,0} \,.
\label{ems.18.2}
\en
This is because the difference between the trajectories is
\be
|\phi_{R,0}| \left| \frac{|s({\bar \tau})|}{\si} - 1\right| = |\phi_{R,0}| \left| \sqrt{1+{\bar \tau}^2/4\si^4} - 1\right| \leqslant  |\phi_{R,0}| \frac{\bar \tau}{2\si^2} \ll u {\bar \tau} \,.
\en
So we approximately get classical motion and hence the full Bohmian result.

Using
\be
S = -\frac{1}{2} \tan^{-1}\left( \frac{\bar \tau}{2\si^2} \right) + u \left( \vp_R - \frac{1}{2}u{\bar \tau}\right) - \frac{( \vp_R - u{\bar \tau})^2}{8\si^2 |s({\bar \tau})|^2} \,,
\en
the classical equation for the scale factor becomes
\be
\al'^2 = u^2+ \frac{1}{2|s({\bar \tau})|^2}  + \frac{\phi_{R,0} u {\bar \tau}   }{ 2\si^3 |s({\bar \tau})|} + \frac{\phi^2_{R,0}}{4\si^4}\left(\frac{{\bar \tau}^2}{4 \si^2 |s({\bar \tau})|^2} - 1 \right)\,. 
\label{ems.19}
\en 
We have that
\begin{multline}
\left| \frac{1}{2|s({\bar \tau})|^2}  + \frac{\phi_{R,0} u {\bar \tau}   }{ 2\si^3 |s({\bar \tau})|} + \frac{\phi^2_{R,0}}{4\si^4}\left(\frac{{\bar \tau}^2}{4 \si^2 |s({\bar \tau})|^2} - 1 \right) \right| \\
\le \frac{1}{2|s({\bar \tau})|^2}  + \frac{u \left| \phi_{R,0}  {\bar \tau}   \right| }{ 2\si^3 |s({\bar \tau})|} + \frac{\phi^2_{R,0}}{4\si^4}\left|\left(\frac{{\bar \tau}^2}{4 \si^2 |s({\bar \tau})|^2} - 1 \right) \right|
\le \frac{1}{2\si^2} + \frac{u|\phi_{R,0}|}{\si^2} + \frac{\phi^2_{R,0}}{4\si^4} \,,
\label{ems.20}
\end{multline}
where we used $|{\bar \tau}|/2\si |s({\bar \tau})| \leqslant 1$ for the last two terms in order to obtain the last inequality. Using the assumptions that $1 \ll \si u$ and $|\phi_{R,0}|  \lesssim \si$,  we find that 
\be
\al'^2 \approx u^2 \,.
\label{ems.21}
\en
So similarly as in the case of the usual semi-classical approximation (just with the extra condition on the initial value $\phi_{R,0}$), we obtain the full quantum results. For the wave function $\psi_L$ similar results hold.

Consider now the superposition  $\chi= (\chi_R + \chi_L)/\sqrt{2}$. This superposition corresponds to two Gaussian packets that move across each other. Initially and finally they are approximately non-overlapping.{\footnote{The latter statement follows from the fact that the spread of the wave function, which equals $|s({\bar \tau})|=\si\sqrt{1+{\bar \tau}^2/4\si^4}$, goes like ${\bar \tau}/2\si$ for large time, which is much smaller than the distance between the centers of the Gaussians, which equals $(u+v){\bar \tau}$.}} This means that before and after the wave packets cross, the motions of $\phi$ and $\al$ are approximately classically, in agreement with the full Bohmian results, since they will be determined by either $\chi_R$ or $\chi_L$. This is unlike the usual semi-classical approximation. 

A trajectory for the scalar field starting from the left (i.e., $\phi < 0$) will end up on the left, while a trajectory starting on the right will end up on the right. The reason is that because of equivariance of the distribution $|\chi|^2$, the $|\chi|^2$-probability to start from the left equals the probability to end up on the left (which equals $1/2$). Since trajectories do not cross (since the dynamics is given by a first-order differential equation in time), the probability for trajectories to start on the left (right) and end up on the right (left) must be zero. This means that for all trajectories starting on the left, we have a transition from $\al = u \phi$ to $\al = -v \phi$ and the opposite transition for trajectories starting on the right. This is unlike the full Bohmian analysis where trajectories exist where such a transition does not occur. 

In conclusion, we see that the Bohmian semi-classical approximation is in better agreement with the exact Bohmian results than the usual semi-classical approximation. We came to this conclusion by making the comparison on the level of the actual trajectories for $\al$ and $\phi$. We did not attempt to make a comparison in the context of standard quantum theory, since the standard quantum interpretation of the Wheeler-DeWitt equation is problematic due to the problem of time. But it is clear that for approaches to quantum theory that would associate (approximately) the classical evolutions \eqref{ems.6.01} and \eqref{ems.7.02} to the superposition $\Psi_R+\Psi_L$ (like perhaps the Consistent Histories or Many Worlds theory) the usual semi-classical approximation would fare worse than the Bohmian semi-classical approximation.

\section{Conclusion}
We have shown how semi-classical approximations can be developed using Bohmian mechanics. We have obtained these approximations from the full Bohmian theory by assuming certain degrees of freedom to evolve approximately classically. This was illustrated for non-relativistic systems. If there is a gauge symmetry, like in electrodynamics or gravity, then extra care is required in order to obtain a consistent semi-classical theory. By eliminating the gauge symmetry (either by imposing a gauge or by working with gauge-independent degrees of freedom), we were able to find a semi-classical approximation in the case of scalar quantum electrodynamics. For quantum gravity, eliminating the gauge symmetry (more precisely the spatial diffeomorphism invariance) is notoriously hard. We have only considered the simplified mini-superspace approach to quantum gravity, which describes an isotropic and homogeneous universe, and where the diffeomorphism invariance is explicitly eliminated. More general cases in quantum gravity still need to be studied. For example, for the case of inflation theory, where one usually considers quantum fluctuations on a classical isotropic and homogeneous universe, it should not be too difficult to develop a Bohmian semi-classical approximation. 

Apart from possible applications in quantum cosmology, such as inflation theory, it might also be interesting to consider potential applications in quantum electrodynamics or quantum optics. In particular, since the results may be compared to the predictions of full quantum theory, this could give us a handle on where to expect better results for the Bohmian semi-classical approximation compared to the usual one in the case of quantum gravity where the full quantum theory is not known. That is, it might give us better insight in which effects are truly quantum and which effect are merely artifacts of the approximation.

Further developments may include higher order corrections to the semi-classical approximation. One way of doing this might be by following the ideas presented in \cite{norsen10,norsen15}. As explained there, one might introduce extra wave functions for a subsystem in addition to the conditional wave function. These wave functions interact with each other and the Bohmian configuration. By including more of those wave functions one presumably obtains better approximations to the full quantum result.

Finally, although we regard the Bohmian semi-classical approximation for quantum gravity as an approximation to some deeper quantum theory for gravity, one could also entertain the possibility that it is a fundamental theory on its own. At least, there is presumably as yet no experimental evidence against it (unlike the usual semi-classical approximation which is ruled out as fundamental theory by the experiment of Page and Geilker).

\section{Acknowledgments}
It is a pleasure to thank Detlef D\"urr, Sheldon Goldstein, Christian Maes, Travis Norsen, Nelson Pinto-Neto, Daniel Sudarsky and Hans Westman for useful discussions and comments. This work was done partly at the University of Liege, with support from the Actions de Recherches Concert\'ees (ARC) of the Belgium Wallonia-Brussels Federation under contract No.\ 12-17/02, at the LMU, Munich, with support of the Deutsche Forschungsgemeinschaft. Currently the support is by the Research Foundation Flanders (Fonds Wetenschappelijk Onderzoek, FWO), Grant No. G066918N.

\end{document}